\documentclass[twocolumn]{aastex631}

\usepackage{amsmath}
\usepackage{graphicx}
\usepackage{xcolor}
\usepackage{scrextend}
\usepackage{CJKutf8}

\providecommand{\sorthelp}[1]{}

\shorttitle{background systematics in low surface brightness imaging}
\shortauthors{Liu et al.}
\graphicspath{{./}{figures/}}

\begin{document}

\title{A Recipe for Unbiased Background Modeling in Deep Wide-Field Astronomical Images}

\correspondingauthor{Qing Liu}
\email{qliu@astro.utoronto.ca}

\author[0000-0002-7490-5991]{Qing Liu \begin{CJK}{UTF8}{gbsn}(刘青)\end{CJK}}
\affil{David A. Dunlap Department of Astronomy \& Astrophysics, University of Toronto, 50 St. George St., Toronto, ON M5S 3H4, Canada}
\affil{Dunlap Institute for Astronomy and Astrophysics, University of Toronto, Toronto ON, M5S 3H4, Canada}

\author[0000-0002-4542-921X]{Roberto Abraham}
\affil{David A. Dunlap Department of Astronomy \& Astrophysics, University of Toronto, 50 St. George St., Toronto, ON M5S 3H4, Canada}
\affil{Dunlap Institute for Astronomy and Astrophysics, University of Toronto, Toronto ON, M5S 3H4, Canada}

\author[0000-0002-5236-3896]{Peter G. Martin}
\affil{Canadian Institute for Theoretical Astrophysics, University of Toronto, 60 St. George St., Toronto,
ON M5S 3H8, Canada}

\author[0000-0003-4381-5245]{William P. Bowman}
\author[0000-0002-8282-9888]{Pieter van Dokkum}
\affil{Department of Astronomy, Yale University, New Haven, CT 06520, USA}

\author[0000-0003-0327-3322]{Steven R. Janssens}
\affil{Centre for Astrophysics and Supercomputing, Swinburne University, Hawthorn VIC 3122, Australia}

\author[0000-0002-4175-3047]{Seery Chen}
\affil{David A. Dunlap Department of Astronomy \& Astrophysics, University of Toronto, 50 St. George St., Toronto, ON M5S 3H4, Canada}
\affil{Dunlap Institute for Astronomy and Astrophysics, University of Toronto, Toronto ON, M5S 3H4, Canada}

\author[0000-0002-7743-2501]{Michael A. Keim}
\affil{Department of Astronomy, Yale University, New Haven, CT 06520, USA}

\author[0000-0002-2406-7344]{Deborah Lokhorst}
\affiliation{NRC Herzberg Astronomy \& Astrophysics Research Centre,
5071 West Saanich Road, Victoria, BC V9E2E7, Canada}

\author[0000-0002-7075-9931]{Imad Pasha}
\author[0000-0002-5120-1684]{Zili Shen}
\affil{Department of Astronomy, Yale University, New Haven, CT 06520, USA}

\author[0000-0001-5310-4186 ]{Jielai Zhang \begin{CJK}{UTF8}{gbsn} (张洁莱)\end{CJK}}
\affil{Centre for Astrophysics and Supercomputing, Swinburne University, Hawthorn VIC 3122, Australia}
\affil{Australian Research Council Centre of Excellence for Gravitational Wave Discovery (OzGrav), Australia}



\begin{abstract}
Unbiased sky background modeling is crucial for the analysis of deep wide-field images, but it remains a major challenge in low surface brightness astronomy. Traditional image processing algorithms are often designed to produce artificially flat backgrounds, erasing astrophysically meaningful structures. In this paper, we present three ideas that can be combined to produce wide-field astronomical data that preserve accurate representations of the background sky: (1) Use of all-sky infrared/sub-mm data to remove the large-scale time-varying components while leaving the scattered light from Galactic cirrus intact, with the assumptions of (a) the underlying background has little power on small scales, and (b) the Galactic cirrus in the field is optically thin on large scales; (2) Censoring of frames contaminated by anomalously prominent wings in the wide-angle point-spread function; and (3) Incorporation of spatial covariance in image stacking that controls the local background consistency. We demonstrate these methods using example datasets obtained with the Dragonfly Telephoto Array, but these general techniques are prospective to be applied to improve sky models in data obtained from other wide-field imaging surveys, including those from the upcoming {\em Vera Rubin Telescope}.

\end{abstract}



\section{Introduction} \label{sec:intro}

Investigation of low surface brightness phenomena in the Universe has evolved recently into one of the most active areas of astrophysics (e.g., \citealt{2010AJ....140..962M}, \citealt{2015MNRAS.446..120D}, \citealt{2015ApJ...807L...2K} \citealt{2015ApJ...800L...3W}, \citealt{2016MNRAS.456.1359F}, \citealt{2016ApJ...823..123T}, \citealt{2017ApJ...834...16M}, \citealt{2018ApJ...857..104G}, \citealt{2020ApJ...894..119D}, \citealt{2021ApJS..256....2D}, \citealt{2022ApJ...933...47C}, \citealt{2022MNRAS.513.1459M}). Remarkably, much of this progress comes on the heels of trailblazing work done almost five decades ago by astronomers working in the era of photographic plates (e.g. \citealt{1971ApJ...163..195A}, \citealt{1971ApJ...169L...3W}, \citealt{1974AJ.....79..671K}, \citealt{1976AJ.....81..954S}). Following these seminal papers, progress slowed down as astronomers transitioned to electronic sensors, which (at least historically) have been difficult to use for low surface brightness imaging, although over the intervening decades important studies were still being undertaken with small CCDs (e.g. \citealt{1997AJ....114..635D}, \citealt{1997ApJ...482..659D}). 

The limiting factor in low surface brightness imaging is usually not the aperture of the telescope, but rather, the level of control over sources of systematic bias. The observer must accurately measure light from extended sources embedded in a spatially and temporally varying sky background, in the presence of internal reflections, scattering, and stray light from sources within and outside the field of view (e.g., \citealt{2009PASP..121.1267S}, \citealt{2014PASP..126...55A}, \citealt{2015MNRAS.446..120D}, \citealt{2019arXiv190909456M}, \citealt{2022A&A...657A..92E}). However, the payoff for investigations that meet these challenges is proving to be enormous. Examples of important
extragalactic science papers based on low surface brightness observations include
studies of nearby dwarf galaxies (e.g. \citealt{2012AJ....144....4M}, \citealt{2015ApJ...807...50B}, \citealt{2017ApJ...847....4G}, \citealt{2017MNRAS.467.2019R}, \citealt{2018ApJ...856...69D}, \citealt{2018ApJ...860...65M}, \citealt{2021ApJS..256....2D}, \citealt{2022ApJ...933...47C}), ultra-diffuse galaxies 
(UDGs; e.g, \citealt{2015ApJ...798L..45V}, \citealt{2015ApJ...809L..21M}, \citealt{2016ApJS..225...11Y}, \citealt{2018Natur.555..629V}), and galaxy outskirts
(e.g., \citealt{2009AJ....138.1417T}, \citealt{2016ApJ...830...62M}, \citealt{2018ApJ...855...78Z}, \citealt{2022ApJ...932...44G}, \citealt{2022MNRAS.515.5335L}). Non-extragalactic science cases also abound, including investigations of solar system objects (e.g., \citealt{2009Natur.461.1098V}), debris disks (e.g., \citealt{2014AJ....148...59S}, \citealt{2020AJ....159...31H}), light echoes (e.g., \citealt{2008ApJ...681L..81R}, \citealt{2010A&A...519A...7O}) and Galactic cirrus (e.g., \citealt{2013ApJ...767...80I}, \citealt{2016A&A...593A...4M}, \citealt{2020A&A...644A..42R}). Below a surface brightness level of 29--30 mag/arcsec$^{2}$, it is both predicted by theory and indicated by observation that there exists a wealth of new low surface brightness structures awaiting exploration by ongoing and future deep wide-field surveys.


One particularly challenging source of systematic bias in wide-field low surface brightness imaging is determination of the two-dimensional `sky background'.  Conventionally speaking, components of this background include light originating above the atmosphere (unresolved extragalactic sources, zodiacal light, diffuse Galactic cirrus), light originating in the atmosphere (airglow emission lines, atmospheric cirrus), light from the telescope (internal reflections, scattering from roughness in optical surfaces and imperfect baffling), and data reduction artifacts (flat-fielding errors, low-quality modeling of electronic bias structure and drift). The key point is that a perfect pipeline should preserve some elements of the large-scale emission while eliminating others, depending on the observer's science goals. In general, one wishes to remove the components that are time-variable, and keep the time-invariant components intact.\footnote{This is similar to, but not quite the same as, saying one wishes to preserve the emission above the atmosphere and remove the rest. One such subtle exception is the removal of zodiacal light, which originates above the atmosphere though its intensity is a function of time (depending mainly on Earth's orbital position).} Throughout the work, we will refer to the \textbf{large-scale time-varying signal pattern} in an astronomical image as the `background' that one aims to remove in the data reduction\footnote{This definition would need modification in cases where the diffuse emission of interest \textit{does} have temporal evolution, {\em e.g.} light echoes. However, in general, the astronomically interesting low surface brightness emission does not vary with time.}  

In this paper, we present multiple approaches to modeling the sky background in deep images: (1) principled background modeling {with the inclusion of dust templates based on Planck observations}, (2) the incorporation of information from the wide-angle point-spread function (PSF) in imaging observations to filter out frames with anomalous background scattered light, and (3) a method for incorporating spatial correlations in low surface brightness emission as a tool for improving the fidelity of local background estimates. These tools can be combined to allow astrophysically interesting emission in the image to be preserved, while eliminating/controlling contaminant emission/systematics from the atmosphere and telescope. 

This work is in a series on the subject of controlling background systematics in low surface brightness imaging. 
An outline of this paper follows. In Section \ref{sec:cirrus}, we discuss the challenges in background systematics, in particular, the Galactic cirrus emission. We describe the data used to illustrate the methods in Section~\ref{sec:data}. Section~\ref{sec:sky_model} presents the background modeling for individual frames using the Planck dust model. Section~\ref{sec:wide_PSF} demonstrates the method of controlling the consistency of the wide-angle PSF for a collection of frames. Section~\ref{sec:GP_model} presents the method of controlling local background variations in the process of image stacking based on the spatial correlation of pixels. The three methods described in Sec.~\ref{sec:sky_model},  Sec.~\ref{sec:wide_PSF}, and Sec.~\ref{sec:GP_model} are combined to optimally reduce our data, and the result is presented in Section~\ref{sec:mosaic}. Section~\ref{sec:discussion} discusses the following: (1) how deep wide-field imaging surveys can benefit from our techniques and Dragonfly data products, and (2) the physical backgrounds and motivation of dust tracers used in Sec.~\ref{sec:sky_model}. Finally, Section~\ref{sec:summary} summarizes our work.


\section{Challenges to obtaining unbiased Sky Background Estimates}
\label{sec:cirrus}

Improper treatment of the sky background is the primary factor limiting high precision photometry of faint extended sources. Because this is well known in the community, there are numerous studies in which authors take special care to model the background, with varying degrees of success (e.g., \citealt{2015ApJS..220....1A}, \citealt{2015ApJ...800L...3W}, \citealt{2016MNRAS.456.1359F}, \citealt{2017ApJ...834...16M}, \citealt{2018PASJ...70S...4A}, \citealt{2018ApJ...857..104G}, \citealt{2020ApJ...894..119D}, \citealt{2022A&A...657A..92E}, \citealt{2022ApJ...932...44G}, \citealt{2023MNRAS.520.2484K}). 
The most common approach is to use a 2D background model obtained from a box-averaged estimator moving across the image (e.g., the approach taken by \texttt{SExtractor}; \citealt{1996A&AS..117..393B}). This method is appropriate for {sources with angular sizes that are small relative to the box size}, but often leads to over-subtraction of extended emission on scales comparable to (or exceeding) the box size chosen to remove the background variation. In this case, astrophysically interesting low surface brightness structures with sizes comparable to the box size could be mistakenly treated as part of the sky background.\footnote{We emphasize that the appropriateness of any particular approach to sky modeling depends on the science goals of the research. The main element in choosing any particular approach is the need for precision at low surface brightness levels, and the main technical drivers are the angular size of the target relative to the sky footprint of the sensor. 
The examples in this paper are not intended to argue that any particular sky modeling approach is `wrong', but rather to highlight limitations that may (or may not) matter, depending on the science goals.} 

\begin{figure*}[!htbp]
    \centering
    \includegraphics[width=0.65\textwidth]{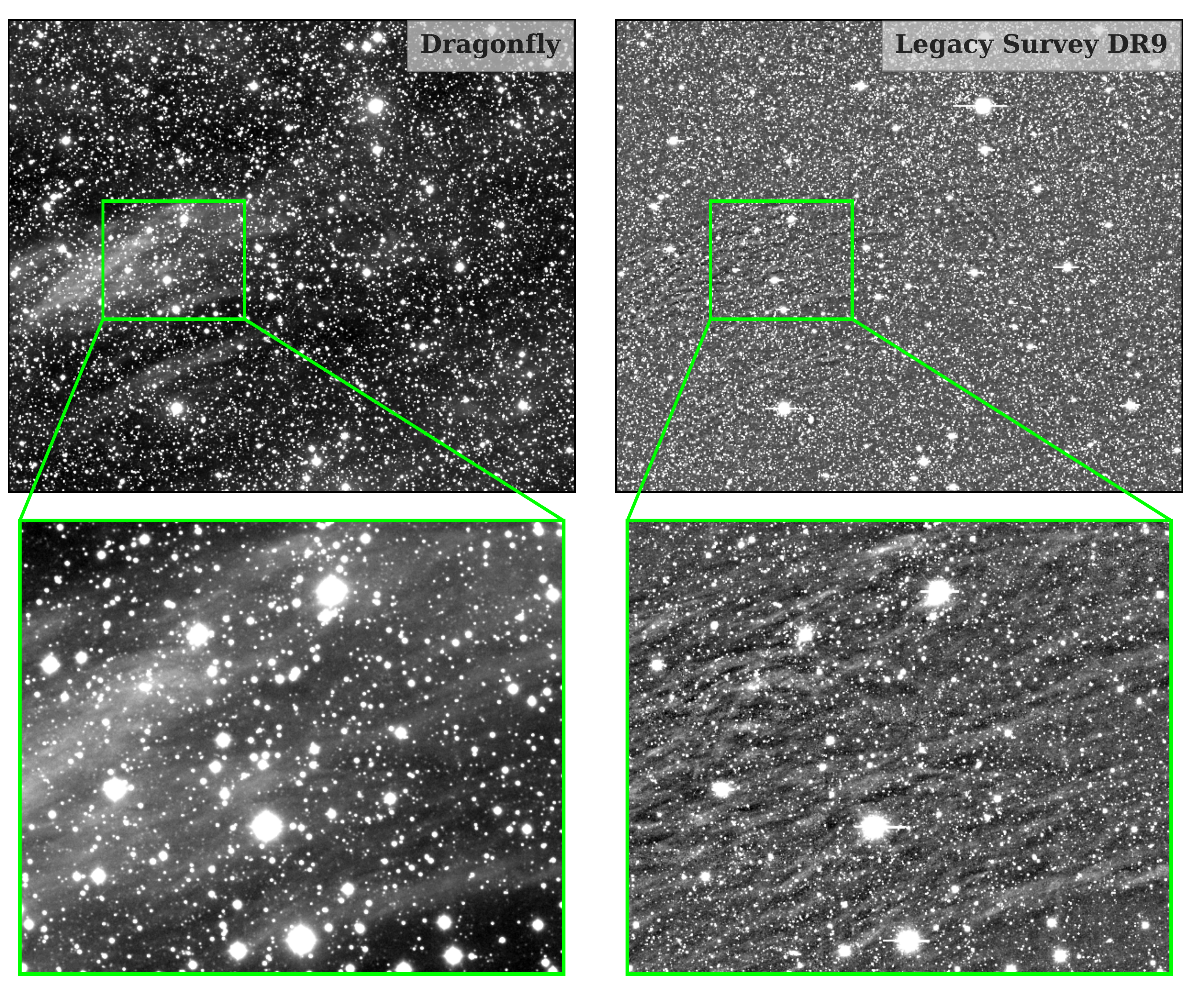}
    \caption{Reduced images of the same sky area taken by Dragonfly (left) and DECam (right) in \textit{r} band. The DECam image is retrieved from the DESI Legacy Imaging survey DR9. The field-of-view (FoV) is 2.5$^\circ$ x 2.1$^\circ$. The zoom-in FoV is 0.6$^\circ$ x 0.5$^\circ$. {The image scales used for the display are from 1\% to 95\% quantiles of pixel values.} The figure illustrates that large-scale low surface brightness structures (in this case, the Galactic cirrus) can be dramatically suppressed due to background subtraction.}
    \label{fig:cirrus_df_decals_panoramic}
\end{figure*}

An example is displayed in Figure~\ref{fig:cirrus_df_decals_panoramic}, which presents a comparison of images obtained with the Dark Energy Camera (DECam) and the Dragonfly Telephoto Array (hereafter Dragonfly, the properties of which are described in Appendix \ref{sec:telescope}). The DECam image is retrieved from Data Release 9 of the DESI Legacy Imaging survey\footnote{https://www.legacysurvey.org/} (\citealt{2019AJ....157..168D}). Because DECam's sensor is a moscaic CCD array, the limiting scale for sky background modeling per detector is the CCD chip size of $9\times18$ arcmin. {The sky subtraction in the \mbox{{\sl legacypipe} pipeline}} runs 2D spline fitting on box estimates with a box size of $\sim$4.5 arcmin (\citealt{2019AJ....157..168D}), which effectively high-pass filters the image on this scale. Dragonfly's monolithic sensor design and background treatment preserve large-scale low surface brightness structures, at the cost of having much poorer resolution than DECam.\footnote{The Vera Rubin Telescope camera has a mosaic of 189 sensors (each with a chip size of 13.5 arcmin) to provide a 3.5-degree field-of-view with high resolution. Background modeling challenges for its camera are expected to be similar to those faced by DECam.} 

More sophisticated background modeling techniques can do a better job of distinguishing the various components of the sky background on an image. {One approach that has turned out to be very successful is to} decompose the sky into different components using forward modeling.
For example, a realistic background model composed of multiple sources (zodiacal light, interstellar medium, and extragalactic background) has been demonstrated in \cite{2022A&A...657A..92E}, using the NASA/IPAC Infrared Science Archive (IRSA) background calculator, supplemented by a ray-traced stray light component. These authors demonstrate the efficacy of calibrating Euclid Visual instrument (VIS) data from simulations, with radiometric calibration provided by a combination of sky flats and calibration maps. These techniques will be used in the Euclid/VIS pipeline, expected to make a marked improvement in the photometry of low surface brightness emission as a result. \cite{2022ApJ...925..219L} presents a forward modeling method that simultaneously models the scattered light from the extended PSF wings of bright stars in the field together with a large-scale sky pattern. Even with conventional background modeling software, dedicated treatments and modification of algorithms are necessary; for example, recently, \cite{2023MNRAS.520.2484K} has shown that a combination of \texttt{SExtractor} with novel source masking techniques can improve the accuracy of sky estimates around faint sources obtained with this software tool in simulated data.


The most subtle emission component in deep wide-field images (and the ultimate limiting factor in very sensitive low surface brightness surveys) is optical Galactic cirrus emission. Visible wavelength `emission' from Galactic cirrus originates from the \textit{scattering} of starlight by interstellar dust grains in the Milky Way. The IR/sub-mm light of Galactic cirrus, however, originates from \textit{thermal emission}. Space-based far-infrared and sub-millimetre telescopes, such as the IR Astronomical Satellite (IRAS) mission, the Herschel Space Observatory, and the Planck Satellite, have been used to identify and analyze cirrus at IR/sub-mm wavelengths (e.g., \citealt{1984ApJ...278L..19L}, \citealt{boulanger1988}, \citealt{1989ApJ...346..773G}, \citealt{1991ApJ...376..335P}, \citealt{boulanger1996}, \citealt{2010ApJ...713..959V}, \citealt{2011MNRAS.412.1151B}, \citealt{planck2011-7.0}, \citealt{planck2011-7.12}, \citealt{planck2013-XVII}, \citealt{planck2013-p06b}, \citealt{planck2014-XXIX}, \citealt{2017A&A...597A.130B}, \citealt{planck2016-l11B}). Emission of the gas coupled to this dust is seen via 21cm radio emission (e.g. \citealt{boulanger1996}, \citealt{2003A&A...411..109M}, \citealt{2015ApJ...809..153M}, \citealt{2017ApJ...834..126B}), which will be revealed more by Square Kilometre Array (SKA; \citealt{2009IEEEP..97.1482D}) and the Next Generation Very Large Array (ngVLA; \citealt{2018ASPC..517...15S}). Due to the low temperature of the dust grains, they are most prominent at far-infrared/sub-mm wavelengths. However, deep imaging surveys probing below 28 mag/arcsec$^{2}$ in the optical bands have revealed the presence of Galactic cirrus across the sky (e.g., \citealt{2015MNRAS.446..120D}, \citealt{2016MNRAS.456.1359F}, 
\citealt{2017ApJ...834...16M}, \citealt{2019ApJS..240....1Z}, \citealt{2020A&A...644A..42R}, \citealt{2021ApJS..257...60Z}, \citealt{2022arXiv220316545L}). 

With the increasing sensitivity of wide-field surveys in recent years, Galactic cirrus is now known to be a major limiting factor in low surface brightness observations. A troublesome aspect of this component is that substructures, such as tidal features and stellar streams, resemble cirrus in shape, brightness, and (sometimes) color, and thus the detection and measurement of faint galactic substructures is hampered by lack of knowledge of the distribution of cirrus on large scales. Examples also include investigations of extended galactic halos and dwarf satellites, which can overlap with cirrus patches. 
Problems such as these will only become more pronounced as imaging surveys push to lower and lower surface brightness levels, e.g, those to be undertaken by Euclid and the Vera Rubin Telescope (\citealt{2022A&A...657A..92E}, \citealt{2022MNRAS.513.1459M}).  

One person's trash is another's treasure. The Galactic cirrus `contaminant' in extragalactic science is of great interest to researchers interested in the interstellar dust at visible wavelengths in the Milky Way, particularly those seeking to constrain their physical properties and conditions (e.g., \citealt{2013ApJ...767...80I}, \citealt{2016A&A...593A...4M}, \citealt{2021MNRAS.508.5825M}, \citealt{2022arXiv220316545L}, \citealt{2023MNRAS.519.4735S}, \citealt{2023ApJ...948....4Z}). Historically, using Galactic cirrus complexes at visible wavelengths as diagnostics has been difficult because of the low surface brightness nature of the structures and their large angular sizes, which can span many degrees on the sky. Precise measurement of the Galactic cirrus emission is a huge challenge for background models applied to CCD images, which mostly apply stacking strategies to data obtained with mosaiced sensors. Background models in conventional pipelines can be biased by these low surface brightness structures, leading to over-subtraction of the signal at an early stage in the analysis, as shown in Figure~\ref{fig:cirrus_df_decals_panoramic}. A survey intended to map Galactic cirrus with high radiometric precision at visible wavelengths would require exquisite control over large-scale systematics in order to preserve these structures. Such control over the sky background is now achievable because of technical advances, as described below. 


\section{Example Datasets} \label{sec:data}

Our example images were chosen to show background patterns typical of the kind encountered in wide-field astronomical images. One field is centered on a prominent Galactic cirrus complex, while the other field shows diffuse cirrus contamination more representative of the kind seen on high galactic latitude images. The data were obtained by the Dragonfly Telephoto Array (Appendix \ref{sec:telescope}), with basic data reduction steps described in Appendix \ref{sec:reduction}. Readers interested in additional details are referred to \cite{2020ApJ...894..119D}. 

\subsection{Field 1: The Spider Field (Extreme Cirrus Emission)} \label{sec:Spider_target}

The `Spider' field is a region of (mostly) optically thin Galactic cirrus centering at $(l,b)\sim(135^\circ, 40^\circ)$, with a distance of $\sim$320 pc (\citealt{2020A&A...633A..51Z}, \citealt{2023ApJ...942...70M}). Dragonfly observations of the full Spider field were taken in 2021. These data are comprised of six tiled fields and one central field, centered at $(\alpha,\delta) = (10{\colon}37{\colon}30.0, +73{\colon}28{\colon}45.0)$. Located at a height z = 205 pc above the Galactic Plane, the Spider field is at the top of the arch of the North Celestial Pole Loop (\citealt{2015ApJ...809..153M}, \citealt{2017ApJ...834..126B}, \citealt{2022ApJ...937...81T}, \citealt{2023ApJ...942...70M}). The central region of the Spider field was previously observed by Dragonfly in 2014 and 2017, with earlier configurations of the array (\citealt{2023ApJ...948....4Z}). In \cite{2023ApJ...948....4Z}, the Dragonfly Spider data is investigated jointly with Herschel data to test dust and scattering models. 

\subsection{Field 2: A High-galactic Latitude Field Affected by Cirrus Background} \label{sec:uw_target}

The other example field, UW1787, is drawn from the Dragonfly Ultrawide (UW) survey\footnote{The Dragonfly UW survey is an ongoing low surface brightness photometric survey undertaken by Dragonfly. It will cover the entire imaging footprint of SDSS ($\approx$10,000 square degrees) with $\sim$2500 pointings. The typical 1$\sigma$ depth of Dragonfly UW survey is 28.5 mag/arcsec$^2$ on 10$\arcsec$×10$\arcsec$ scales in \textit{g}-band. A future paper will describe the survey strategy, science goals, and data quality assessment in more detail. The majority of the UW fields are free from cirrus or only embeds partial area of cirrus, however, a small fraction of them suffer more significantly from high galactic latitude cirrus background.}. The field center is $(\alpha,\delta) =  (22{\colon}26{\colon}47.4, -06{\colon}33{\colon}32.5)$, which corresponds to $(l,b)=(57.2^\circ, -50.0^\circ)$. UW1787 presents a wide area of faint, diffuse cirrus emission in the west of the field. An initial inspection of Planck dust maps at the same sky position shows that the diffuse light in the background of the Dragonfly data corresponds to regions of dust emission, indicating that the structure does not originate from inadequate sky modeling or camera artifacts. UW1787 serves as an example field from imaging surveys. The field is not specifically targeted at a cirrus complex (such as the Spider field) but is affected by diffuse Galactic cirrus background.

\section{Technique 1: Background Modeling Using a Planck Dust Model} \label{sec:sky_model}
A major development that can be used now to determine astrophysically meaningful backgrounds on CCD images is the public availability of all-sky far-infrared/sub-mm data from the Planck satellite. 
In this section, we make use of the all-sky dust thermal emission model from the Planck 2013 data release. The Planck all-sky thermal dust model is described and presented in \citet{planck2013-p06b} (see also \citealt{planck2016-XLVIII}), which is retrieved from the Planck Legacy Archive\footnote{We use the  HFI\_CompMap\_ThermalDustModel\_2048\_R1.20.fits Healpix map (nside=2048) downloaded from Planck Legacy Archive (\url{http://pla.esac.esa.int}). The single file can also be found at IRSA (\url{https://irsa.ipac.caltech.edu/data/Planck/release_1/all-sky-maps/previews/}).}. We use dust radiance ($\mathcal{R}$) as a tracer for dust emission. Dust radiance provides a good estimate of dust column density in optically thin ($E(B-V)<0.3$) sky areas. Alternatively, we tested using maps of dust optical depth at $\nu_0 = 353$ GHz ($\tau_{353}$), which are noisier than radiance maps, and yield nearly identical results because our sky modeling focuses on large scales. Further discussion on the motivation for the choice of dust tracers in the Planck model based on physical considerations is deferred until Section \ref{sec:discussion_dust}. The dust emission maps are extracted from the Healpix map using the \texttt{reproject\_from\_healpix} function in the \texttt{reproject} Python package. 

\subsection{Issues with Conventional Background Modeling}
\label{sec:conventional_sky}

The first step in building a conventional background model (e.g., with {\tt SExtractor}) is to create a mask that can be used to eliminate compact sources. Then, unmasked pixels are fit by a low-order polynomial to model the large-scale variation in the image:
\begin{equation}
    I(x,y) = B_n(x,y) + \epsilon(x,y)\,,
    \label{eq1:poly}
\end{equation}
where $I(x,y)$ is the pixel intensity map, $B_n(x,y)=\Sigma\,b_{ij} \cdot P_{ij}(x,y)\, (i, j \in [0,n], i+j \leq n)$ is the background model, which is a sum of n\textit{-th} order orthogonal bivariate Legendre basis polynomial functions $P_{ij}$ as a function of pixel coordinates with $b_{ij}$ equal to the amplitudes, and $\epsilon$ is the residual noise. $I(x,y)$ is usually further smoothed/evaluated by a box estimator (e.g., that used by SExtractor), denoted as $\hat{I}(x,y)$, before fitting to remove small-scale background fluctuations.
This sky background model $B_n(x,y)$ is then subtracted from the original image $I(x,y)$.

There are two problems with this procedure. The first is that it can be difficult to construct the mask. This is because popular source detection methods rely on peak finding algorithms, but diffuse structures generally do not have a single, well-defined peak in their light distributions, so they would be counted as part of the sky background and subtracted off from the final image. The second problem is even more fundamental: some sources (such as Galactic cirrus) may be so widely distributed that they are effectively blended into the sky over most of the image, and there may not be enough uncontaminated pixels left to allow an adequate large-scale sky model to be built.  


\begin{figure*}[!htbp]
    \centering
    \includegraphics[width=\textwidth]{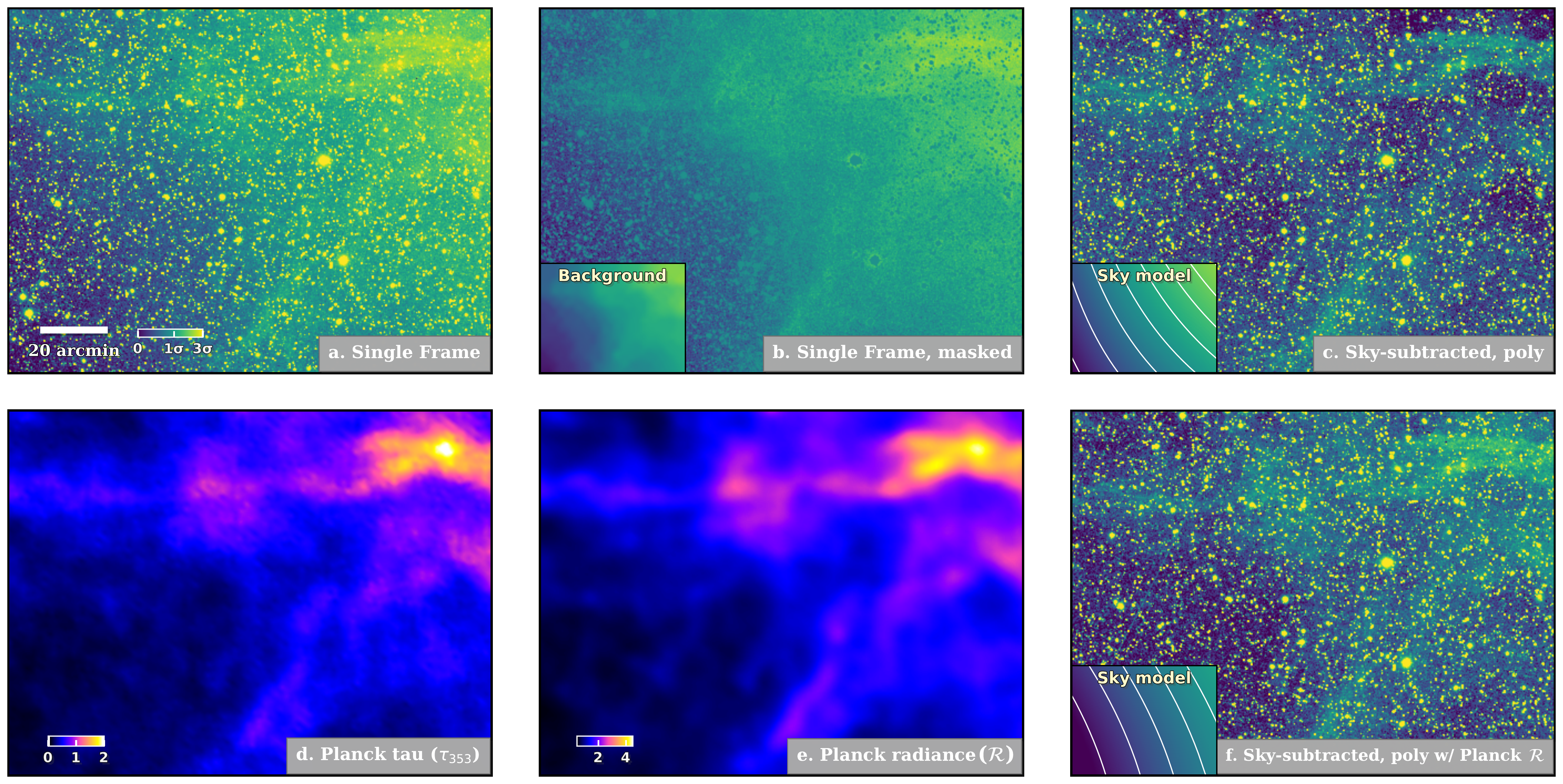}
    \caption{Comparison of sky modeling of an individual frame using a single polynomial sky model and that with Planck dust model. The image is a $2.6^{\circ} \times 2^{\circ}$ 10-min exposure of the Spider field obtained by a single camera of Dragonfly. (a): Image of the dark subtracted, flat-fielded single frame. (b): Image with sources masked, with masked pixels replaced with the median sky values for display. The inset shows the smooth 2D background evaluated by \texttt{SExtractor}. (c): Sky-subtracted image using a single 2nd-order polynomial sky model. The inset shows the sky model. (d): Planck optical depth map at 353 GHz in units of $10^{-5}$. (e): Planck dust radiance map in units of [$10^{-7}{\,}\rm{W{\,}m^{-2}{\,}sr^{-1}}$]. (f): Sky-subtracted image using a 2nd-order polynomial sky model fitted using the techniques described in this paper, {which makes use of the Planck template}. The inset shows the improved sky model. Image scales are stretched to enhance low surface brightness features. Zero-points of panels (c) and (f) are shifted to have the same quantiles.}
    \label{fig:Planck_sky_1}
\end{figure*}

The top three panels of Figure \ref{fig:Planck_sky_1} demonstrate how sky modeling can be biased. The demonstration uses an example exposure of the Spider field obtained by a single camera from the Dragonfly array. Images are stretched with an arcsinh function to enhance the visibility of low surface brightness features. Panel (a) displays a reduced (dark subtracted, flat-fielded) single frame, $I(x,y)$. The masked image shown in Panel (b) was created by masking all detected sources with signal-to-noise ratio (S/N) $>$2 using \texttt{SExtractor} with \texttt{BACK{\_}SIZE} = 128 $\rm pix$ (corresponding to $6\arcmin$). {The masked area was then grown with three iterations of morphological dilation}. With sources masked, a smooth 2D background, $\hat{I}(x,y)$, was evaluated using a [128$\times$128] pix$^2$ box mode estimator, displayed in the inset of panel (b). A second-order (n=2) polynomial model, $B_{2}(x,y)$, was used to fit this smooth background $\hat{I}(x,y)$. This sky model was then subtracted off from the original image. The background-subtracted image is shown in panel (c) with the polynomial sky model $B_{2}(x,y)$ displayed as an inset. 
Limiting the polynomial degree to n=2 aims to preserve the large-scale emission over $1.3\times1$ deg$^2$, representing a relatively conservative sky estimate.

An inspection of panel (b) in Figure~\ref{fig:Planck_sky_1} shows that stars and galaxies are masked, but most cirrus is not. As a result, the fitted polynomial sky model is strongly biased by cirrus. For example, the cirrus signal on the upper right corner of the image is significantly suppressed, even though it is almost certainly real (see below). On the other hand, the left half of the image presents residual light from inadequate sky subtraction.
Fitting the polynomial model on $\hat{I}(x,y)$ estimated with a larger box estimator [256$\times$256] yields a similar sky model in this case, and no choice of box size yields an appropriate model throughout the whole image. 
The sky area close to edges and corners is particularly problematic because convolution-based box estimators are not good at capturing sky variations near the sides of images.

\subsection{Dust Priors: A Better Path for Background Modeling?}


Our approach to background modeling improves upon the procedure described in the previous section by incorporating ancillary data, such as information on dust emission revealed by satellites such as Planck and Herschel (see similar considerations in \citealt{2015MNRAS.446..120D} and \citealt{2017ApJ...834...16M}). The key assumption here is that the optical dust emission (from scattering) is correlated with infrared radiance (derived from thermal emission), \textit{at least on large scales}. This assumption probably holds in cases where the dust grains are in thermal equilibrium with the interstellar radiation field (ISRF), so a given population of dust grains has a fixed spectral energy distribution (SED) that can be well described by a modified blackbody (MBB) law (\citealt{planck2013-p06b}). The exact shape of the SED depends on many parameters (e.g., grain size distribution, chemical composition, porosity, optical depth; \citealt{2011piim.book.....D}). On the scale of our sky modeling ($\sim$ 1 deg), we assume dust grains nearby are well mixed, and their ensemble properties are similar. Panel~(d) and panel~(e) of Fig.~\ref{fig:Planck_sky_1} show the registered Planck optical depth ($\tau_{353}$) map and Planck dust radiance ($\mathcal{R}$) map, respectively. A comparison with Panel~(c) shows that strong similarity exists between the sub-mm dust map and the optical data. This reinforces our assumption that the same dust population emits thermally in the sub-mm and scatters the starlight in the optical (as expected by dust models, \citealt{2023ApJ...948....4Z}, and see discussion in Sec. \ref{sec:discussion_dust}). This correlation is also supported by observations in various other studies (e.g., \citealt{2013ApJ...767...80I}, \citealt{2016A&A...593A...4M},  \citealt{2020A&A...644A..42R}, \citealt{2021MNRAS.508.5825M}, \citealt{2022arXiv220316545L}, \citealt{2023MNRAS.519.4735S}). Note that the resolution of Planck (beam FWHM $\sim5\arcmin$) is about a factor of 50 worse than that of Dragonfly, so we can only apply our approach to removing background on large angular scales. Correlations between optical and IR radiance are likely to be poorer on small scales in any case, due to factors such as measurement uncertainties, variations in the optical depths, size distributions, and dust compositions (\citealt{2003ApJ...598.1017D}). We assume that these higher-order effects are negligible in large-scale sky modeling.

To turn these ideas into a practical method for sky modeling, we first downsample the optical image by reprojecting it to a coarser grid using cubic splines to degrade the resolution for the match with Planck maps. This downsampling process does not affect large-scale background patterns. $\mathcal{R}$ is used as the tracer of dust emission $I_{\rm{dust}}$. A low-order polynomial dust model $B_{n,\,\rm{dust}}(x,y)$ is then fitted on the Planck dust emission map $I_{\rm{dust}}(x,y)$:
\begin{equation}
    I_{\rm{dust}}(x,y) = B_{n,\,\rm{dust}}(x,y) + \epsilon_{\rm{dust}}(x,y)\,.
\end{equation}
{The fitting parameters are the polynomial coefficients for the dust model $b_{ij,\rm{dust}}$ ($i, j \in [0,3]$, $i+j \leq 3$)}. This low-order polynomial model $B_{n,\,\rm{dust}}(x,y)$ ({Planck template}) is scaled and added as an extra {component} (i.e., a `regularized' term) to the polynomial fitting in Equation \ref{eq1:poly} when fitting the smooth 2D background evaluated from the optical image, $\hat{I}(x,y)$, with a new sky model $B_{\rm{n,\,{sky}}}(x,y)$:
\begin{equation}
    \hat{I}(x,y) = B_{\rm{n,\,{sky}}}(x,y) + \lambda \cdot B_{n,\,\rm{dust}}(x,y) + \epsilon(x,y)\,.
\end{equation}
We choose $n_{\rm{sky}}=2$ and $n_{\rm{dust}}=3$ to account for only large-scale variations in the sky background and the dust components. Free parameters include the amplitudes of the polynomial coefficients of the `intrinsic' sky $b_{ij,\,\rm{sky}}$ ($i, j \in [0,2]$, $i+j \leq 2$) with $b_{00,\,\rm{sky}}$ equal to the mean sky brightness and the scaling factor $\lambda$. Limiting the dust model $B_{3,\,\rm{dust}}(x,y)$ to a higher order than the sky model $B_{2,\,\rm{sky}}(x,y)$ empirically assumes that {the underlying sky pattern has little power in its brightness variation on small scales, }
which appears to be the case in both the Spider field (a cirrus-rich field) and UW1787 (a field with diffuse cirrus).

Panel~(f) of Fig.~\ref{fig:Planck_sky_1} shows the resulting sky-subtracted image. The zeropoint is shifted to the same quantile as that of panel (c) for comparison. The improved second-order polynomial sky model (the `effective' sky without the dust component), $B_{\rm{2,\,{sky}}}(x,y)$, is shown in the inset, which represents the inferred underlying large-scale sky background {after incorporating the Planck template}. Comparing panel~(f) to panel~(c) reveals a clear difference in the sky background of the two images. On the whole, the overall emission pattern in panel~(f) is a much closer match to the Planck data. In particular, panel~(f) preserves significant cirrus emission known to exist from the Planck data that is suppressed by the original sky modeling.  {Although it can be argued that this match is a necessary consequence of employing Planck templates in sky modeling, it is worth noting that the sky is modeled only over degree scales while the spatial correspondence exists down to much smaller scales. As we will show in Section \ref{sec:mosaic}, this approach is further validated by the close match to FIR data, which is independent of Planck models.}

\begin{figure*}[!htbp]
    \centering
    \includegraphics[width=\textwidth]{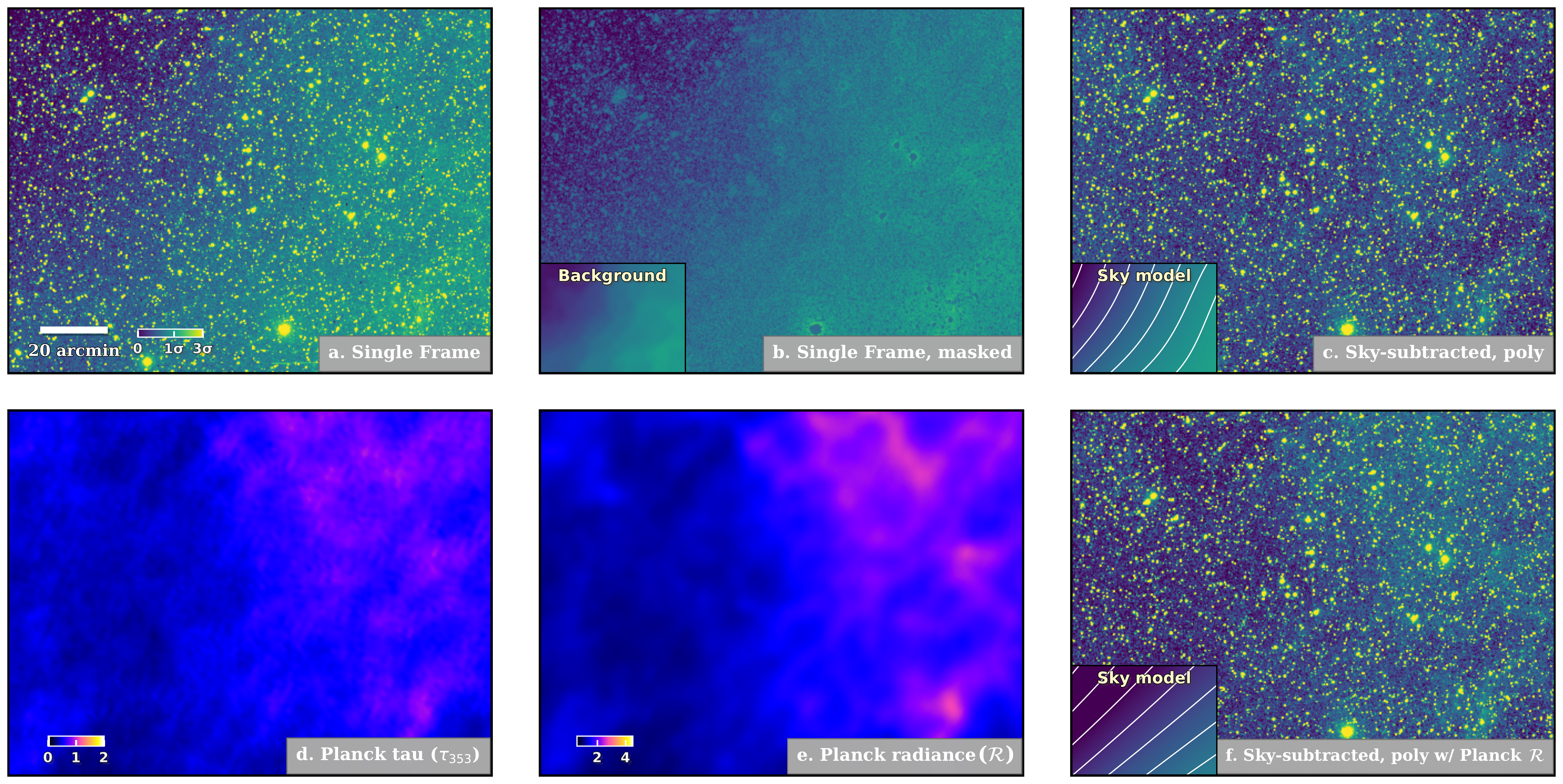}
    \caption{Comparison of sky modeling of an individual frame using a single polynomial model and that with Planck dust model. The image is a $2.6^{\circ} \times 2^{\circ}$ 10-min exposure from a single camera of Dragonfly. The layout is the same as Fig.~\ref{fig:Planck_sky_1} but shows a frame of UW1787, a typical survey field affected by diffuse high-galactic latitude cirrus emission. Image scales are stretched to enhance low surface brightness features.}
    \label{fig:Planck_sky_uw}
\end{figure*} 

The morphology of the Galactic cirrus in the Spider field is complicated. However, the general approach described above turns out to capture the range of structures seen, {with further cases shown in Appendix \ref{sec:sky_model_spider_fig}}. 

The above example shows the application on fields with extreme cirrus emission. It is interesting to investigate the outcome of the technique applied to typical survey fields affected by high-galactic latitude cirrus background. Figure~\ref{fig:Planck_sky_uw} shows the technique applied on UW1787, a field drawn from the ongoing Dragonfly UW survey. Although no compact cirrus structures are present in the image, the benefit from the improved sky modeling is still significant. The original sky model in panel~(c) tends to flatten the image by smearing out the cirrus emission. Comparing panel~(c) and panel~(f) shows clear improvements in the upper right corner and near the left edge of the field.

To summarize, introducing dust priors {by including templates based on FIR/sub-mm observations} improves estimates of the sky background on deep wide-field images in the presence of Galactic cirrus emission. The key advantage of using Planck data in sky modeling is that it has all-sky coverage and is well-calibrated in its systematics (\citealt{planck2013-p06b}). Background modeling using IR/sub-mm survey priors is a promising approach that can be applied to wide-field surveys affected by Galactic cirrus. However, limitations in the CCD sensor size and the large beam size of IR/sub-mm data remain challenges for a direct application to data obtained with large telescopes using mosaic CCD arrays (see discussion in Section \ref{sec:discussion_surveys}), which requires further investigations and improvements.



\section{Technique 2: Wide-angle PSF Control} \label{sec:wide_PSF}

The extended wing of the point-spread function, sometimes referred to as the wide-angle PSF or the stellar aureole, is one of the major sources of systematics in low surface brightness imaging (e.g., \citealt{2014A&A...567A..97S}, \citealt{2020MNRAS.491.5317I}, \citealt{2022ApJ...925..219L}). Light from the PSF wings of stars and galaxies can permeate the entire FoV at low surface brightness levels. {At low surface brightness levels, e.g., by 29.5 mag/arcsec$^2$ for the Burrell Schmidt Telescope (a telescope optimized for low surface brightness imaging)}, it becomes indistinguishable from the sky background (\citealt{2009PASP..121.1267S}), complicating both the measurement of the PSF itself and the photometry of faint, diffuse sources such as dwarf satellites, UDGs, and extended galactic disks/halos. The origin of the wide-angle PSF is still debated, and it may arise from a variety of sources, some of which are from the telescope or within instruments (e.g., \citealt{1996PASP..108..699R}, \citealt{2009PASP..121.1267S}), while others are in the atmosphere, including thin atmospheric cirri, aerosols, and dust (e.g., \citealt{2013JGRD..118.5679D}). The fact that some of the wide-angle PSF originates within telescopes and/or instruments is the motivation for the design of low surface imaging optimized telescopes such as the Dragonfly Telephoto Array.


\subsection{Issues with Blind Image Stacking} \label{sec:temporal_wide_PSF}

An under-appreciated aspect of the wide-angle PSF is that it is not static. 
This variability leads to several complications. For example, there could be bias if a previously measured wide-angle PSF has a marked difference from the PSF in actual data. Second, systematics can occur when stacking exposures that are taken at different times. As different PSFs collapse into one in the final coadd, the details of low surface brightness features are also blurred, which reduces the sensitivity of the surface brightness levels. To achieve precision photometry at low surface brightness levels, the temporal variation of the wide-angle PSF needs to be considered (\citealt{2014A&A...567A..97S}, \citealt{2022ApJ...925..219L}).

There is little doubt that much of the temporal variation in the wide-angle PSF of ground-based telescopes originates from the atmosphere, but some component of it originates in the telescope. Most work on this subject has been done in the context of space-based telescopes, for which thermal fluctuations are often the major source of PSF variability (\citealt{2007ApJS..172..203R}). Variation in the wide-angle PSF of ground-based telescopes has only been observationally confirmed in a limited number of studies.\footnote{Note that the PSF variability being considered here occurs at radii far larger than the rapid PSF variability caused by atmospheric turbulence, which has been well-investigated in the context of adaptive optics imaging.} An early investigation by \cite{2002A&A...384..763M} found a difference in the extended PSF wing of the telescope at Haute-Provence Observatory separated by three months, although the cause was still unclear. More recently in \cite{2022ApJ...925..219L}, we showed that one source of the temporal variation in the wide-angle PSF of Dragonfly is lens cleanliness: as dust accumulates on optical surfaces, the scattered light in the wings of the PSF increases.\footnote{{Not} all sources of wide-angle PSF variation are instrumental. Dust accumulation changes the wide-angle PSF wing contribution slowly (typically over a timescale of months) but atmospheric factors operate on shorter timescales. For example, investigation of Dragonfly's PSF suggests that aerosol optical depth may also play a role in shaping the wide-angle PSF. Atmospheric effects vary on a timescale of hours, and even sometimes on sub-hourly timescales. 
The physical origin of temporal variations in the wide-angle PSF is beyond the scope of this paper.} 

\subsection{Detection of Frames with Abnormal Wide-angle PSFs}

\begin{figure*}[!htbp]
\centering
  \resizebox{0.75\hsize}{!}{\includegraphics{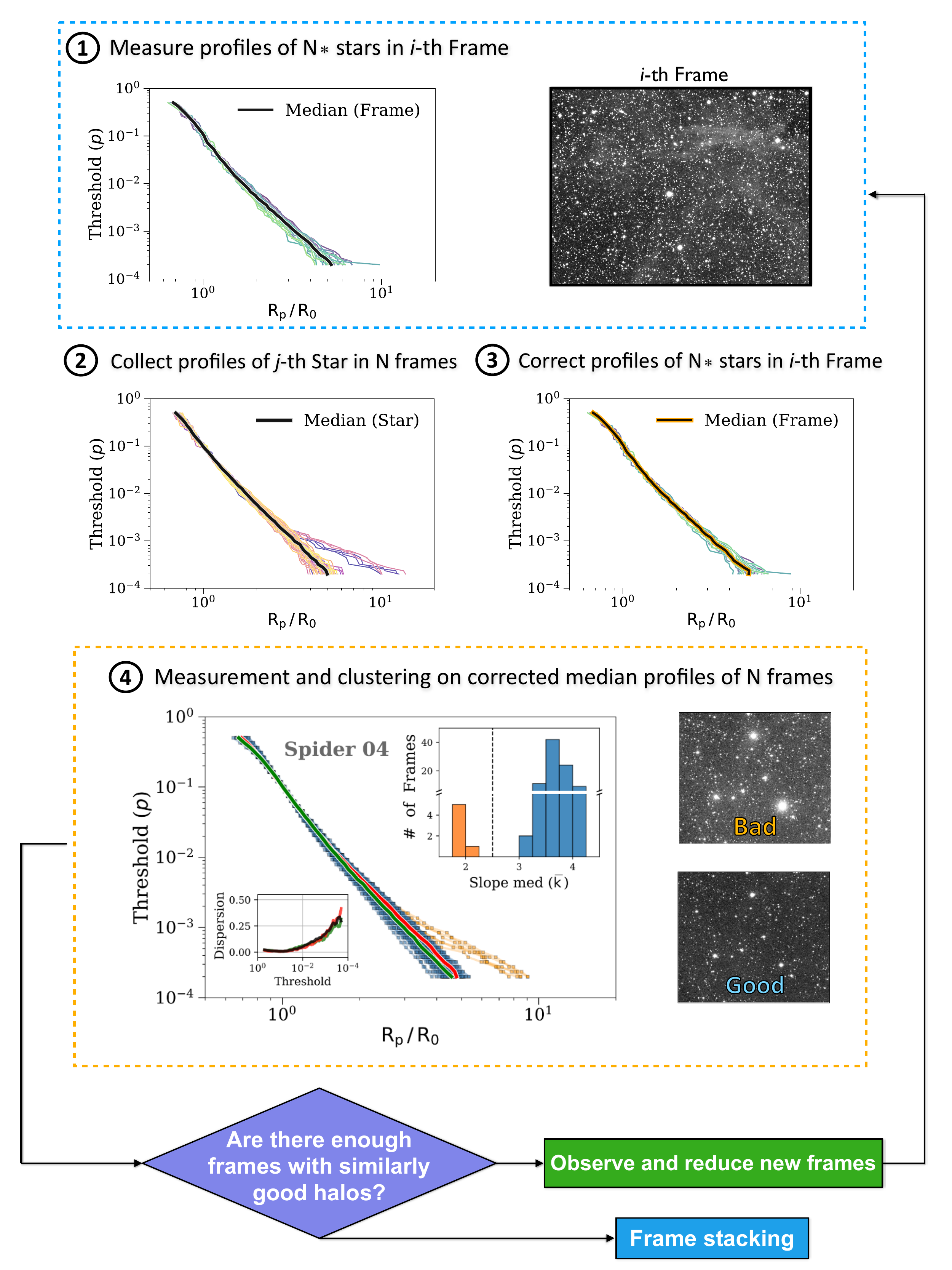}}
  \caption{Example workflow of the wide-angle PSF assessment. From top to bottom: (1) For a given frame, threshold-radii profiles of bright stars are extracted and normalized. The median profile over stars is shown as the black curve. (2) For each star, a median profile over frames in the same filter is acquired. The profiles of an example star in \textit{r}-band frames are displayed as solid curves. The black curve indicates the median stellar profile. (3) The profiles in (1) are divided by the corresponding median profile (the solid curve in (2)) to correct background systematics around individual stars. The profiles are multiplied by the overall mean profile for visualization. The profile averaged over stars after correction is displayed as the orange-black curve. (4) The median corrected threshold-radii profiles, each representing a frame, are collected and run with a clustering algorithm to identify outliers. Slopes and dispersions of profiles are also measured as absolute metrics. See further descriptions in the text and Fig.~\ref{fig:halo_uw1787}. (5) Frames with well-behaved wide-angle PSFs are passed to the stacking if a sufficient number of frames for the requested depth have been acquired. Otherwise, the target can be added back to the observing queue to obtain more frames.}
\label{fig:halo_schematic}
\end{figure*}

\begin{figure}[!htbp]
\centering
  \resizebox{\hsize}{!}{\includegraphics{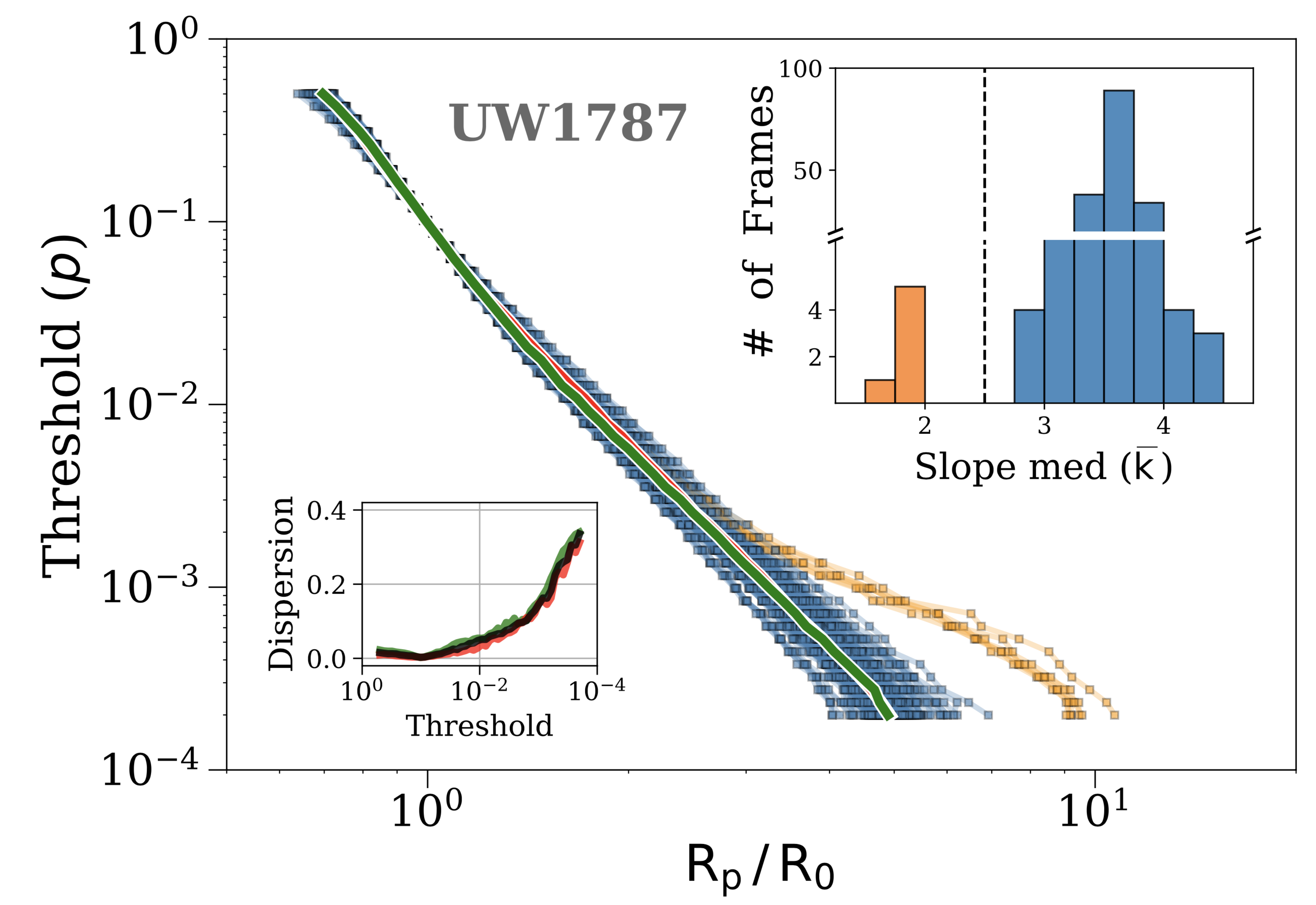}}
  \caption{Wide-angle PSF assessment on frames of UW1787 showing the threshold-radii profiles measured from individual frames. The $x$-axis is the normalized radii at which the brightness level drops to certain fractions ($p$) of the saturation level. Each profile represents the median profiles averaged over bright stars. Outliers identified by the clustering algorithm are displayed in orange. The median \textit{g}/\textit{r} profile is shown by the thick green/red curve. The slopes measured from the median profiles are shown in the top right histogram. The lower left inset panel shows the dispersion of profiles along the threshold.}
\label{fig:halo_uw1787}
\end{figure}


The most straightforward approach to reducing the systematics caused by temporal variations of the wide-angle PSF wings in stacked images is to simply remove bad data frames before they contribute to the final stack.
In \cite{2022ApJ...925..219L}, we presented a technique based on deep wide-field Dragonfly imaging, which simultaneously fits the extended stellar PSF wings and the background on a pixel-based level to characterize the wide-angle PSF via forward modeling. Although in principle one can run the full Bayesian modeling on individual frames to derive the PSF of each frame, the time cost of thoroughly sampling the posterior distribution of the model parameters for a large ($\gtrsim$100) number of frames is high. The presence of bright Galactic cirrus in this example dataset also disturbs the PSF modeling. Therefore, for the purpose of efficiently filtering out bad frames, we do \textit{measurement} instead of \textit{fitting}. In this section, we describe an efficient technique based on measurements of brightness profiles, which can be used to identify frames with anomalously prominent wide-angle PSFs. This approach is based on the measurement of stellar profiles, followed by a clustering analysis (or an assessment based on absolute metrics). The procedure {is applied on sky-subtracted individual frames}, which is described below and illustrated with one of the Spider fields in Fig.~\ref{fig:halo_schematic}. The assessment step is based on panel (4) of Fig.~\ref{fig:halo_schematic}. Figure \ref{fig:halo_uw1787} illustrates the assessment of the UW1787 field.

\subsubsection{Profile Extraction} \label{sec:profile_extraction}

\begin{figure*}[!htbp]
\centering
  \resizebox{0.95\hsize}{!}{\includegraphics{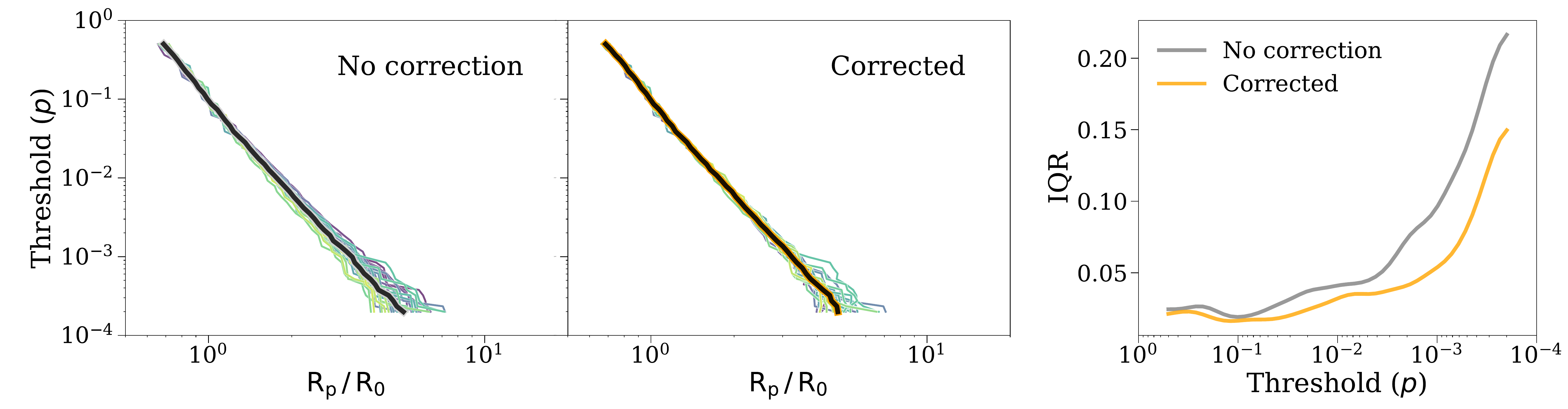}}
  \caption{\textit{Left} and \textit{middle}: threshold-radii profiles extracted from a single frame with and without the systematic correction. The thick solid curves indicate the median profile used as the representative profile of the frame. \textit{Right}: interquartile range measured along the threshold in log space with (orange) and without (gray) correction. The systematic correction largely reduces the dispersion of the profiles due to local background systematics.}
\label{fig:halo_IQR}
\end{figure*}

First, sources detected by \texttt{SExtractor} are cross-matched with the ATLAS catalog (\citealt{2018ApJ...867..105T}) in the two Dragonfly bands ($g$ and $r$). The ATLAS catalog is an all-sky catalog with high completeness and accuracy at the bright end. For the example datasets, we select stars between 8 and 10.5 mag, and use these as bright reference stars (the magnitude range can be tweaked per frame). These stars are strongly saturated at their cores. For each star, we measure its radial averaged surface brightness profile with nearby sources masked. {The radial profile is measured with a list of annuli by taking the 3$\sigma$-clipped mean of unmasked pixels within the annulus. Note the masking procedure can induce potential bias on the measurement in crowded fields (\citealt{2022ApJ...925..219L}, \citealt{2023MNRAS.520.2484K}) where the extended PSF wings of nearby bright sources may `leak' out of the masks.} If more than 80\% of pixels are masked, the star is removed from the sample. Stars close to the field edges and corners are also excluded. The smooth 2D background estimated by \texttt{SExtractor} is used for determining the local background value. {The effect of radial asymmetry of the PSF (e.g., drift, diffraction spikes) is negligible for Dragonfly PSF at large radial angles (\citealt{2022ApJ...925..219L}) and does not affect the assessment below based on relative comparisons.}

Next, instead of using surface brightness profiles directly, we adopt relative measurements based on the saturation level, since the saturation in the cores of stars occurs at roughly the same brightness level in the image, regardless of the local background. For each star, we measure a series of radii, $R_p$, at which the surface brightness drops to a certain fraction, $p$, of the saturation level. Stellar profiles differ due to differences in their magnitudes and positions in the field. To normalize the profiles by the brightness of stars, we fit a linear model in log space for each profile and shift it by $R_0$, the value of the fitted model at a fixed threshold $p_0$. Here we choose $p_0=0.1$ for normalization. The fitting for the linear model is performed inside a range in which the profiles have high S/N and are mostly parallel (corresponding to the range of the first power component of the wide-angle PSF in \cite{2022ApJ...925..219L}). For the example datasets, $p=0.5$ to $0.01$ is chosen. At small fractions or large radii, the profiles begin to divert from each other due to various local background systematics (e.g., Galactic cirrus, scattered light from other stars, inadequate sky modeling). Indeed, these systematics are the primary culprits that make wide-angle PSF assessment difficult in low surface brightness imaging. However, whether being flattened (by external light from other sources) or steepened (by overestimated background), the profiles of \textit{the same} bright star measured among all different frames should be deformed by the same amount, if and only if the wide-angle PSF is consistent. 

Two schemes of discriminating the goodness of the wide-angle PSF are thus feasible: one is to group profiles by stars and compare group properties based on statistical inference, and the other is to correct the systematics among stars for each profile and then compare the relative difference of average profiles from different frames. We adopt the latter scheme and correct the systematics to the first order. 

To illustrate these considerations, the shifted threshold-radii profiles of a bright star sample are shown in panel (1) of Fig.~\ref{fig:halo_schematic}, with different colors indicating different stars. The profile shows how scattered light distributes as a function of radius in the context of classical annuli-based photometry. A representative profile is obtained by taking the median value of the shifted profiles at different thresholds, shown as the black solid curve. Individual profiles suffer from background systematics, although this bias is mitigated in median profiles. 

To correct the bias for each star, the profiles of the same star measured from all frames are collected, as shown in panel (2). A median profile (for each band) is used as the representative of the star. This average stellar profile is used to divide the profile of individual measurements for the corresponding star and in the corresponding band to correct the bias to the first order. Panel (3) shows the corrected profiles of stars where the dispersion of profiles is considerably reduced. This is further illustrated in Fig.~\ref{fig:halo_IQR}. The left and middle panels show stellar profiles extracted from a single exposure, before and after local background correction, respectively. The right panel shows the inter-quartile range (IQR) along thresholds, illustrating how the dispersion is significantly reduced after the correction. 

After correction, each median profile is considered to be representative of the frame. For a given frame, the shape of the median profile is determined by the effective wide-angle PSF per frame. If the wide-angle PSF of a single frame has significantly brighter halos than others, the profile will in general flatten at the outskirts, deviating from the rest of the frames. 
Frames with similarly good wide-angle PSFs will present similar profiles with small dispersion. The median profiles from all frames of the Spider04 field are shown in the main panel of panel~(4) in Fig.~\ref{fig:halo_schematic}. Fig.~\ref{fig:halo_uw1787} shows the median profiles extracted from the UW1787 field. For illustration purposes, the profiles are divided by the median profile of all frames and then multiplied by the median profiles in the corresponding filter.

\subsubsection{Profile Clustering} \label{sec:profile_clustering}

Given the expected behavior of the profiles, we can use an unsupervised clustering method to classify groups of profiles with similarities and identify outliers. Here we choose the DBSCAN algorithm (\citealt{ester1996densitybased}), which classifies samples based on their densities in the feature space and naturally incorporates outlier detection. We use the DBSCAN implementation in the Python package \texttt{scikit-learn}. In DBSCAN, there are two hyperparameters: $N_{min}$ and \texttt{eps}. $N_{min}$ sets the minimum number of samples for neighboring samples to be considered as a cluster. We choose $N_{min}$ to be 5$\%$ of the total frames. The other parameter, \texttt{eps}, is more critical in that it controls the intra-cluster dispersion, i.e., the sparsity of a cluster. Higher \texttt{eps} rejects more samples and classifies them as noise or outliers. The optimal value of \texttt{eps} is specific to the dataset. We run the clustering over a grid of \texttt{eps} and compute silhouette coefficients to evaluate the performance of the clustering. The silhouette coefficient is a score between -1 for improper clustering and 1 for highly dense clustering. Models with the same performance yield equal silhouette coefficients. The best clustering model is chosen as the one with the highest silhouette coefficient. The classifications of group member frames and outliers are indicated by the color coding in panel (4) of Fig.~\ref{fig:halo_schematic} and Fig.~\ref{fig:halo_uw1787}. In panel (4) of Fig.~\ref{fig:halo_schematic}, {example} cutouts of an outlier and a group member are shown as the images on the right.


During this step, we also measure several quantities that characterize the profile. One useful diagnostic is the dispersion of profiles, which quantifies the intragroup distance. The lower left inset of panel (4) of Fig.~\ref{fig:halo_schematic} and that in Fig.~\ref{fig:halo_uw1787} show the dispersion among the median profiles along the threshold (green/red in \textit{g}/\textit{r}-band and black in total). A group associated with unusually large dispersion is likely composed of frames covering a variety of wide-angle PSFs. Such PSF variability could happen when the frames are gathered from a wide time period, or a period with very unstable weather conditions, e.g., a period affected by monsoons. As a result, the effective wide-angle PSF in the coadd image embeds certain smearing effects from image stacking, as noted in Section~\ref{sec:temporal_wide_PSF}. We caution that when combining frames from a variety of observing and telescope conditions, it is important to be aware of such systematics. The use of absolute metrics is discussed further in the next section.

The idea of PSF classification with clustering algorithm is not new. A recent study by \cite{2018MNRAS.478.5671W} developed a method using self-organizing maps combined with principal component analysis to classify the PSF obtained at different times and by different telescopes. We note that the objective and application of the analysis in \cite{2018MNRAS.478.5671W} differ from ours in a few ways. Firstly, the part of the PSF that we aim to characterize is the wide-angle component that extends much further than the range affected by atmospheric turbulence ({\citealt{1996PASP..108..699R}}). Secondly, the profiles clustered in our analysis are the average brightness profiles derived from a single frame, which can be biased compared with that from the intrinsic PSF. Finally, we directly cluster the profiles without dimensionality reduction. As mentioned above, the goal of clustering in our approach is to establish a fast algorithm to identify the main group of profiles, evaluate the similarity of the members, and flag the outliers. 

\subsubsection{Assessment based on Absolute Metrics} \label{sec:metrics_wide_psf}

\begin{figure*}[!htbp]
\centering
  \resizebox{0.98\hsize}{!}{\includegraphics{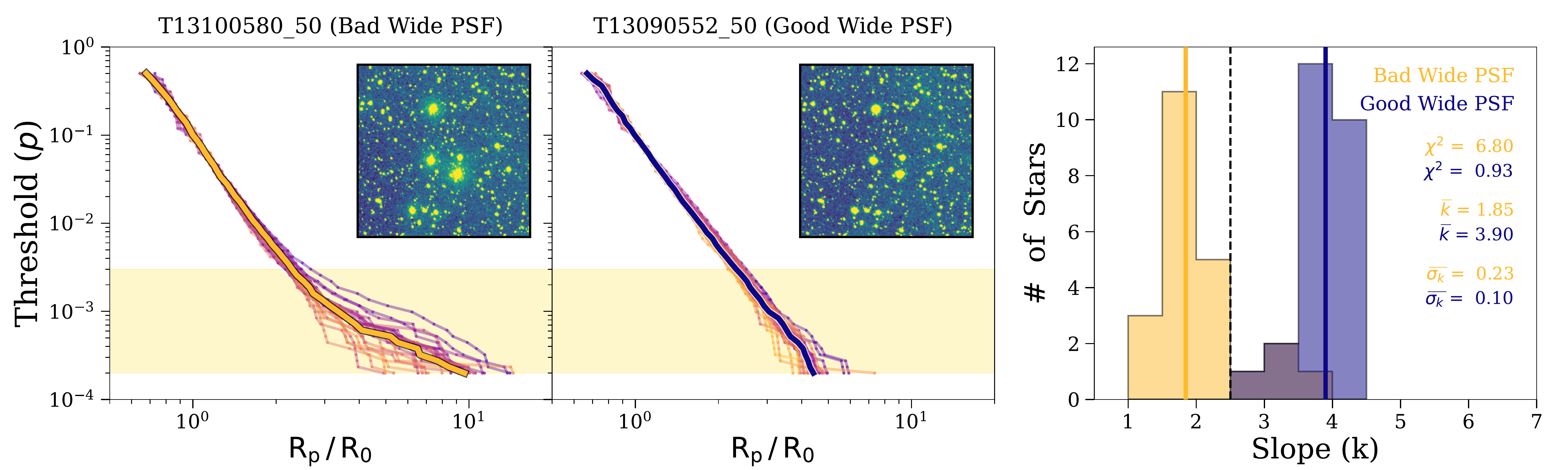}}
  \caption{Wide-angle PSF assessment of two example frames of the Spider field based on absolute metrics. \textit{Left}: threshold-radii profiles of bright stars extracted from a single frame flagged with bad wide-angle PSF. \textit{Middle}: profiles extracted from a single frame passing the assessment. The frames are named by [{\em camera serial number}]$\_$[{\em exposure number}]. The median profiles are indicated by the thick colored curves. {The yellow shaded areas indicate the threshold range in which the slopes are measured.} The postage stamps show 40$\arcmin$ $\times$ 40$\arcmin$ cutouts of the frames. \textit{Right}: histograms of slopes measured from profiles of individual stars of the two frames (orange: bad; blue: good). For each frame, the slope of the median profile is indicated by the vertical colored line and $\bar{k}$. The mean reduced chi-square and the dispersion of slopes of individual profiles are indicated by $\chi^2$ and $\overline{\sigma_k}$. The black dashed line shows a demarcation line (bad/good wide-PSF) based on median slopes.}
\label{fig:halo_frame}
\end{figure*}

In Section~\ref{sec:profile_clustering}, we present the outlier identification based on a clustering algorithm. In general, the clustering has good performance when the majority of the frames present well-behaved wide-angle PSFs with only a small fraction of outliers, or in some rare cases where the profiles form discrete groups resulting from very distinct telescope and/or weather conditions. However, the clustering algorithm assumes at least one well-behaved population and that good frames significantly outnumber bad frames, which is not always true. For example, there are cases where only a small number of frames is obtained in the first place, or where the initial frames have a large dispersion in their wide-angle PSF behaviors. This could happen at the beginning of a long-term sky-scanning survey with each observation separated by a period, or on a wide but relatively shallow survey where the total number of frames is small. Therefore, assessment based on absolute metrics also turns out to be useful for the quality control of the wide-angle PSFs. The final evaluation can use either the clustering output or the absolute metrics, or ideally, a combination of both.

Besides the dispersion of profiles along the threshold shown in the previous section, we also measure the slope of the outskirts of the profile by fitting a linear model between a fixed range at large radii. The slopes contain useful information because the optimal shape of the wide-angle PSF is constrained by several factors, such as the observing site, the weather condition, and the cleanliness of lens surfaces. Therefore, the expected value of the slope has a relatively small variation within measurement tolerance, regardless of which factors degrade the wide-angle PSFs. The slopes are measured on individual profiles of the bright star sample of each frame and on the median profile of the frame in log space. 

Three quantities are useful to characterize and control the wide-angle PSF. One is the slope of the median profile of the frame ($\bar{k}$). Here $\bar{k}=2.5$ is chosen in practice as the discriminative threshold for the example datasets based on empirical runs. The second quantity is the relative standard deviation ($\overline{\sigma_k}=\sigma_k/\bar{k}$) of slopes ($k$) measured from individual stars. There might be edge cases where the average profile presents a reasonable slope within the threshold range, but the with-in frame dispersion is large. Typically, for a frame with stable wide PSF behavior, a reasonable median slope is associated with a small dispersion. We exclude frames with large dispersions $\overline{\sigma_k}>0.3$. Third, a mean reduced chi-square ($\chi^2$) is calculated for each frame on the goodness of fitting linear models on individual bright stars. Profiles with strong deviation from linear models (i.e. large $\chi^2$ values) are usually caused by a prominent extended PSF wing. Frames with $\chi^2$ over three standard deviations above the mean value are excluded. A frame violating any of the three above criteria is flagged with a \texttt{HALO} flag, which stands for excessive scattered light in the outskirts of the wide-angle PSF. Note a correction for background systematics is important to reduce the dispersion of slopes and appropriate $\chi^2$ values.

Figure~\ref{fig:halo_frame} shows profiles of two example frames of the Spider fields obtained by Dragonfly for the wide-angle PSF assessment. The example frames were taken at the same time, but with different cameras. In Fig.~\ref{fig:halo_frame}, the left panel shows profiles extracted from an example frame flagged as \texttt{HALO}, and the middle panel shows that from an example frame that passes the assessment. The median profiles are shown by the thick curves. We choose a threshold from $p=0.0002$ to 0.003 to be the fitting range for the example dataset. The range can be tweaked depending on the surface brightness limit and the magnitude range of bright stars in the field. The slopes measured on individual profiles are shown as histograms in the right panel ofFig.~\ref{fig:halo_frame} . The orange histogram corresponds to frames with the \texttt{HALO} flag, while the blue histogram corresponds to unflagged frames. For each frame, $\chi^2$, $\bar{k}$, and $\overline{\sigma_k}$ are indicated on the right with the corresponding color. {The classification is based on the combination of these metrics. In Appendix \ref{sec:wide_psf_uw_fig}, we show further examples of the wide-angle PSF assessment based on metrics using frames of the UW1787 field.}

\section{Technique 3: Image Consistency Filtering Using Pixel Covariance Constraints}\label{sec:GP_model}


The powerful techniques presented in the previous sections do a good job of preserving astrophysical backgrounds impacted by known sources of contamination, but there are a vast number of ways for anomalies to propagate through pipelines. Therefore, the final ingredient in our recipe for background modeling is to ensure overall consistency in the backgrounds among frames.

\subsection{Issues with Local Background Variation among Frames}

\begin{figure*}[!htbp]
\centering
  \resizebox{0.95\hsize}{!}{\includegraphics{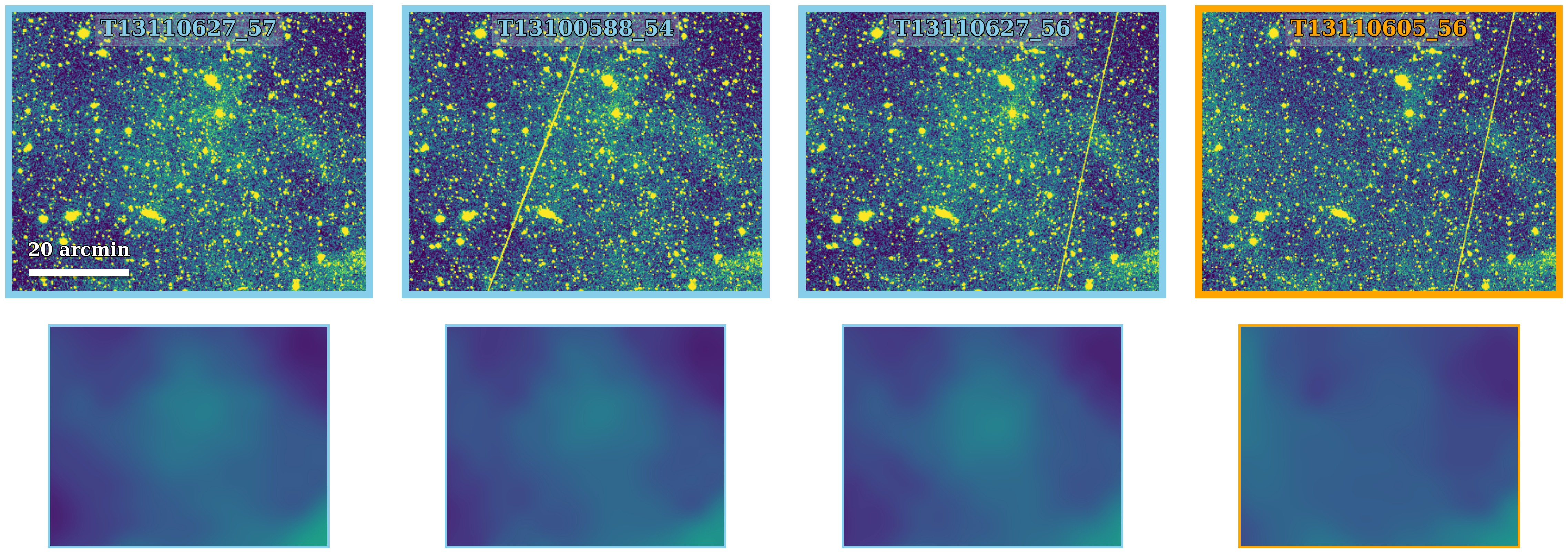}}
  \caption{An example showing local background deviations in individual frames. \textit{Upper row}: the same $1.3^{\circ} \times 1^{\circ}$ cutout areas of four individual frames of the Spider field taken by different cameras or at different times, reprojected into the same grid. Large-scale backgrounds have been subtracted following Sec.~\ref{sec:sky_model}. \textit{Lower row}: background of the upper cutouts evaluated by a [128$\times$128] pix$^2$ box mode estimator. The fourth column shows a problematic frame. Because we use low-order polynomials ($n$=2) in background subtraction (see Sec.~\ref{sec:sky_model}), patterns on scales smaller than $1.3^{\circ} \times 1^{\circ}$ have been preserved and need control. The frames are named by [{\em camera serial number}]$\_$[{\em exposure number}].}
\label{fig:sky_variance}
\end{figure*}

Our approach is best illustrated by example. In Figure~\ref{fig:sky_variance}, the same $1.3^{\circ} \times 1^{\circ}$ cutout regions from four individual frames are shown in the top row. These images were taken from three cameras in the Dragonfly array. Two frames were taken by the same camera, but at different times. It is apparent that the background in the fourth image is anomalous, because it clearly deviates from the others. The inconsistency is even more obvious in the second row, which shows the {smoothed} backgrounds evaluated by a [128$\times$128] pix$^2$ box mode estimator. This example suggests that a useful approach to background modeling is to look for a degree of consistency among frames prior to incorporating them in image stacks. 


A legitimate question one might ask is whether or not cases such as that shown in Fig.~\ref{fig:sky_variance} can be handled by sigma-clipping \footnote{In the context of image stacking, this refers to per-pixel rejection in the combination of registered images. Any pixel with a deviation from the mean/median value greater than a certain threshold (in the unit of Poisson noise) will be rejected.}, a commonly adopted approach to combining images. Unfortunately, the answer is not always `yes', because in low surface brightness imaging, background systematics (such as those from flat-fielding errors) are often at 
levels comparable to that of the Poisson noise in the diffuse emission. 
Discrepancies are not obvious at the level of individual pixels, but they become clear at the level of ensembles of nearby pixels. In other words, the individual pixels in erroneous areas are not significantly deviant -- only the mean in an area is. Bad regions turn out to show strong pixel-to-pixel correlations (or covariance) on individual frames, and they stand out by appearing discrepant when compared to independently obtained images.
These characteristics suggest that other approaches to pixel censoring and stacking might be more successful than sigma-clipping when preservation of background structure is a priority.
The key idea is to incorporate pixel covariance information to make predictions of mean patterns on large scales (the idea of which is similar to the techniques presented in \citealt{2022ApJ...933..155S} but for a purpose of interpolation) and to look for deviations from these patterns as signatures of bad data.

In Sections \ref{sec:GP_outline} and \ref{sec:GP_implementation} we will present an implementation of this idea using Gaussian process regression (GPR), a statistical learning technique often used in machine learning (\citealt{2006gpml.book.....R}). GPR has been applied in a number of studies in many fields of astronomy for non-parametric modeling (e.g., \citealt{2012PhRvD..85l3530S}, \citealt{2017AJ....154..220F}, \citealt{2019ApJ...879..116I}). Other implementations using alternative approaches (such as convolutional neural networks) would likely also work well.


\begin{figure*}[!htbp]
\centering
  \resizebox{\hsize}{!}{\includegraphics{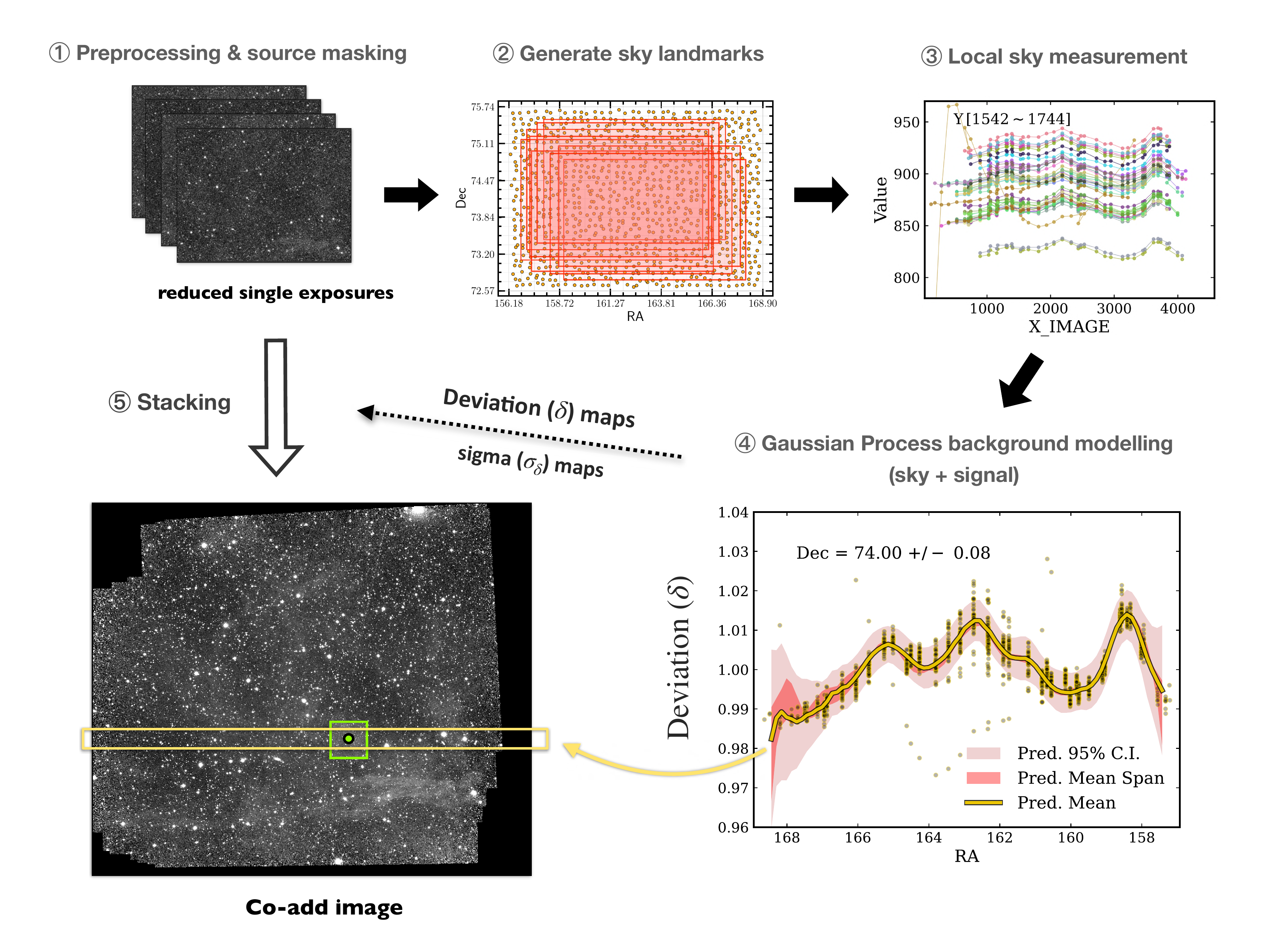}}
  \caption{Workflow schematic of the stacking process with Gaussian process sky modeling. (1) A set of single frames after bias and dark subtraction, flat-fielding, sky subtraction, and wide-angle PSF assessment. (2) A grid of landmarks with random perturbation is generated to cover the FoV. (3) In each frame, the local backgrounds are measured at the positions of landmarks, with sources masked. For illustration, the panel shows measurements from different frames (in different colors) within a slice of the Y-axis. (4) A Gaussian process model is trained using the normalized local sky measurements to predict a mean 2D background pattern and its standard deviation on the RA/dec grid. Deviation maps are generated relative to the predicted mean. For illustration, the 1D projection in a Dec slice is displayed. The deviation is in fractional units. The yellow curve indicates the predicted mean, with the red shaded area showing the 95\% confidence interval. The darker red area shows the span of predictions from five-fold cross-validation. Measurements from individual frames are shown as black dots. The dispersion originates from measurement uncertainties and sky variation within the slice. The gaps between measurements are due to masking.  (5) Single frames are stacked making use of the deviation maps {and the standard deviation maps} generated by (4). The green box illustrates the scale used for measuring the local sky in (3), and the yellow rectangle indicates the Dec slice in (4).}
\label{fig:gp_schematic}
\end{figure*}

\subsection{The Key Idea: Sigma-Clipping on Larger Scales}
\label{sec:GP_outline}


Our approach to modeling backgrounds using GPR will be illustrated using frames from one of the Spider fields, Spider02, which is filled with strong Galactic cirrus emission, and thus, provides a good example\footnote{In fields with Galactic cirrus, the signal is correlated in the sense that it exhibits a smooth power spectrum that is well described by a power law in the Fourier space over a wide range of the electromagnetic spectrum (\citealt{2016A&A...593A...4M}). This will be analyzed in more detail in our follow-up paper.}. In this case, the emission in the background (relative to sources -- note the difference between the term `background' used here and the sky background being subtracted in Sec.~\ref{sec:sky_model}) shows considerable structure, with strong spatial correlations. 

The GPR technique captures the correlation of pixels by using a kernel\footnote{Note that this `kernel' is not the convolution kernel familiar to astronomers. Rather, it refers to the numerical relationship that measures the similarity of data, {\em i.e.} the covariance functions.}. An individual frame, after subtracting the global mean (the mean sky brightness across FoV), is treated as a random draw of this Gaussian process, with the measurement noise mainly coming from Poisson photon statistics. The model predicts a mean pattern combining all frames. Because a Gaussian process by its definition is a stationary process (i.e., {the correlation between two points only depends on their separation vector}), the model implies that the local background at any position does not change. If this is violated, the local background from an individual frame would deviate from the ensemble mean. Averaging out the noise and assuming that the signal is unchanged within the time period of any pair of frames\footnote{This assumption does not hold for some low surface brightness phenomena. For example, light echos in supernova remnants expand at detective rates, revealed by precise differential imaging of observations taken at multiple epochs (\citealt{2008ApJ...681L..81R}). Nevertheless, a temporal background model incorporating a physical model for the evolution of the signal and the cadence of observations is not impossible, and {is worth} exploring in future time-domain low surface brightness astronomy.}, it is natural to infer that the difference originates from the underlying sky, or other residual systematics from data reduction such as flat-fielding errors that need to be controlled. This is an important supplement to the background modeling presented in Sec.~\ref{sec:sky_model} because we use a low-order polynomial to remove the large-scale time-varying pattern, which is not sensitive to background variations at intermediate scales. Here we generate a deviation map for each frame, which is used for masking bad pixel areas or for weighting in the average combination, and then proceed to a sigma-clipped stacking. 

Practical details of our implementation are given in the next subsection. {The} method falls back to pixel-based sigma-clipped combination when pixels have no spatial correlations. When correlations exist and information from a neighborhood of pixels is also used, such stacking methods are very robust, and are particularly useful when the distribution of intensity values at any single pixel is strongly biased or skewed.

\subsection{Implementation}
\label{sec:GP_implementation}

\begin{figure*}[!htbp]
\centering
  \resizebox{0.9\hsize}{!}{\includegraphics{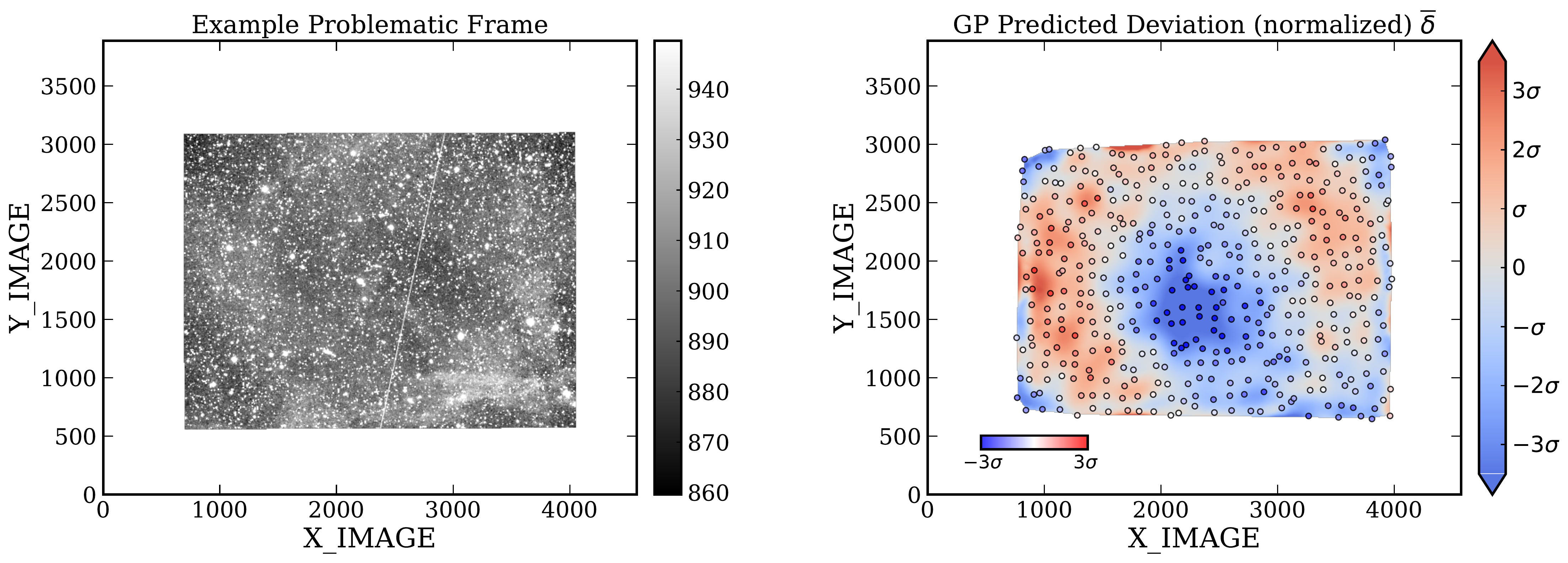}}
  \caption{An example of the sky deviation map (right) generated from the GPR background modeling for a single frame (left) with strong local background deviations. The image scale is stretched to enhance low surface brightness features. {The deviation (from the mean pattern) is in the unit of the standard deviation of the distribution predicted by GPR.} Colored dots represent landmarks at which the local background is measured on a scale of 20$\arcmin$. During image combination, pixel areas with large deviations are rejected or downweighted.}
\label{fig:gp_deviation}
\end{figure*}

The first step is to measure the local background. This is done by using a grid of points on the sky (N$\sim$1000 points sparsely distributed across the entire FoV) with random jitter. We will refer to these points below as ‘landmarks’. The spatial distribution of landmarks in our example data is illustrated in panel (2) of Figure~\ref{fig:gp_schematic}. The key point here is that these landmarks are associated with fixed positions {\em on the sky}. In each frame, we measure the local background value (estimated using the mode) and noise level (rms) within a box centered around each landmark. The choice of box size is not critical, provided it is large enough to typically encompass several nearby landmarks. However, since power at small scales comes mostly from stars, galaxies and sky noise, and at very large scales from the frame limited FoV, 
the box size should not be too small or too large. In our example dataset, we choose a box size of 20 arcmin. Sources detected in the previous \texttt{SExtractor} run are masked. Measurements from regions containing $>$ 80\% masked pixels, including those close to the field edges, are rejected. {The measurements of different frames within a Y-axis slice are shown in panel (3) of Fig. \ref{fig:gp_schematic} with different colors for illustration.}
 
Prior to training, each frame must be normalized to have a common mean and standard deviation. We exclude landmark measurements from individual frames with extreme values that are 10-sigma higher or lower than the global mean, since these are likely contaminated by extremely bright stars or affected by strong camera artifacts such as shutter issues. 
At this point, the measurements from all frames, grouped by filters, are fed into a 5-fold GPR model (a `fold' is an instance of training sample). For cross-validation, in each fold, the GPR model is trained with 80\% of measurements to predict a mean background pattern (sky + signal) and its standard deviation, which vary smoothly across the field. {For illustration, panel (4) of Fig.~\ref{fig:gp_schematic} shows the 1D projection of the predicted mean within a Dec slice as the yellow curve.} The span of the predictions from five folds is shown as the darker red area, indicating the consistency across folds. In this example, the span is very narrow {except around field edges}, but note that the predictions might embed larger dispersion if only a small number of frames are used for training. The mean model is taken from the average of the cross-validation outputs, shown as the gold curve. In our example, we adopt a Radial Basis Fucntion (RBF) kernel with a fixed scale length of $\sim20\arcmin$, matching the box size of the background estimator. Using fixed scale length in GPR modeling allows it to control the spatial scale of the predicted background to which it is sensitive. The scale length can also be leveraged as a fitted parameter in the training process. Noise is included in the training and propagated through the model. The 2-sigma range indicating the 95\% confidence interval of the predicted mean is shown as the red shaded area in panel {(4)} of Fig.~\ref{fig:gp_schematic}. Measurements with abnormal local backgrounds appear as outliers compared to the predicted mean, and these should be omitted from the final stack.

To enhance computation efficiency, we use the GPR model from landmarks to first predict the mean sky + signal value {and its standard deviation} over a sparse grid defined by the landmarks. For each frame, a full image is then interpolated from the sparse grid using a cubic spline {for the mean and sigma}. The areas outside the convex hull of the landmarks are discarded for proper interpolation. Quality control is monitored by measuring the deviation of the target frame relative to the mean model and applying it in image stacking. To do so, we divide each frame by the mean sky + signal model image to generate a deviation map ($\delta$) that indicates the percentage deviation of the local region on a pixel-by-pixel level. {A normalized deviation map ($\overline{\delta}$) is generated by dividing the deviation map by the local standard deviation ($\sigma_{\delta}$) predicted by GPR.}

The final result of applying this process to the fourth problematic frame identified `by eye' in Fig.~\ref{fig:sky_variance} is shown in Fig.~\ref{fig:gp_deviation}.  
An example normalized deviation map derived from the frame showing local background issues is displayed in the right panel of this figure. In this case, the radial symmetry suggests that the anomalous structure was probably caused by flat-fielding errors. 

Deviation maps such as those shown in this figure were used to eliminate bad pixel areas prior to making the final image stacks shown in the next section. 
In this case, pixel areas with deviations higher than {$2\,\sigma_\delta$} were rejected to make the final coadds. In the end, only a small number of problematic frames ($<$ 5\%) showed erroneous local background patterns similar to that in Figure~\ref{fig:gp_deviation}. For the majority of the sky, the background deviation level was under 0.5\%{, which corresponds to the mean $2\,\sigma_\delta$ range. Note the GPR model was loosely constrained by a lack of adjacent landmarks.}

\section{Results: Application to Example Datasets} \label{sec:mosaic}

\begin{figure*}[!htbp]
\centering
  \resizebox{0.65\hsize}{!}{\includegraphics{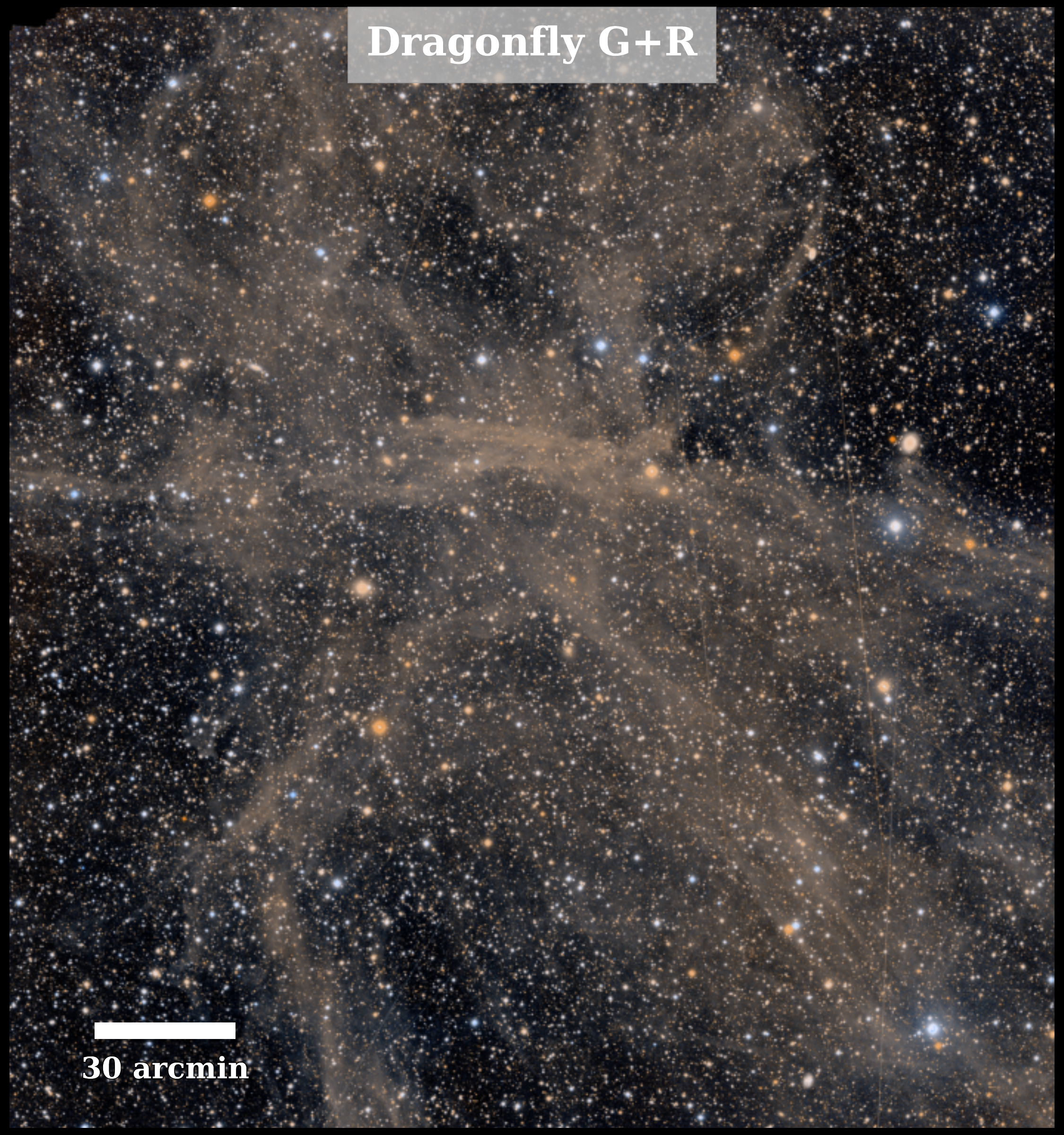}}
  \caption{Dragonfly \textit{g}+\textit{r} mosaic RGB image of the Spider field. The G channel of the RGB is the average of \textit{g} and \textit{r} band data. The mosaic is a tiling of 7 fields obtained across different nights. The field-of-view is $4.2^\circ\times4.5^\circ$.}
\label{fig:Spider_rgb}
\end{figure*}

\begin{figure*}[!htbp]
\centering
  \resizebox{0.7\hsize}{!}{\includegraphics{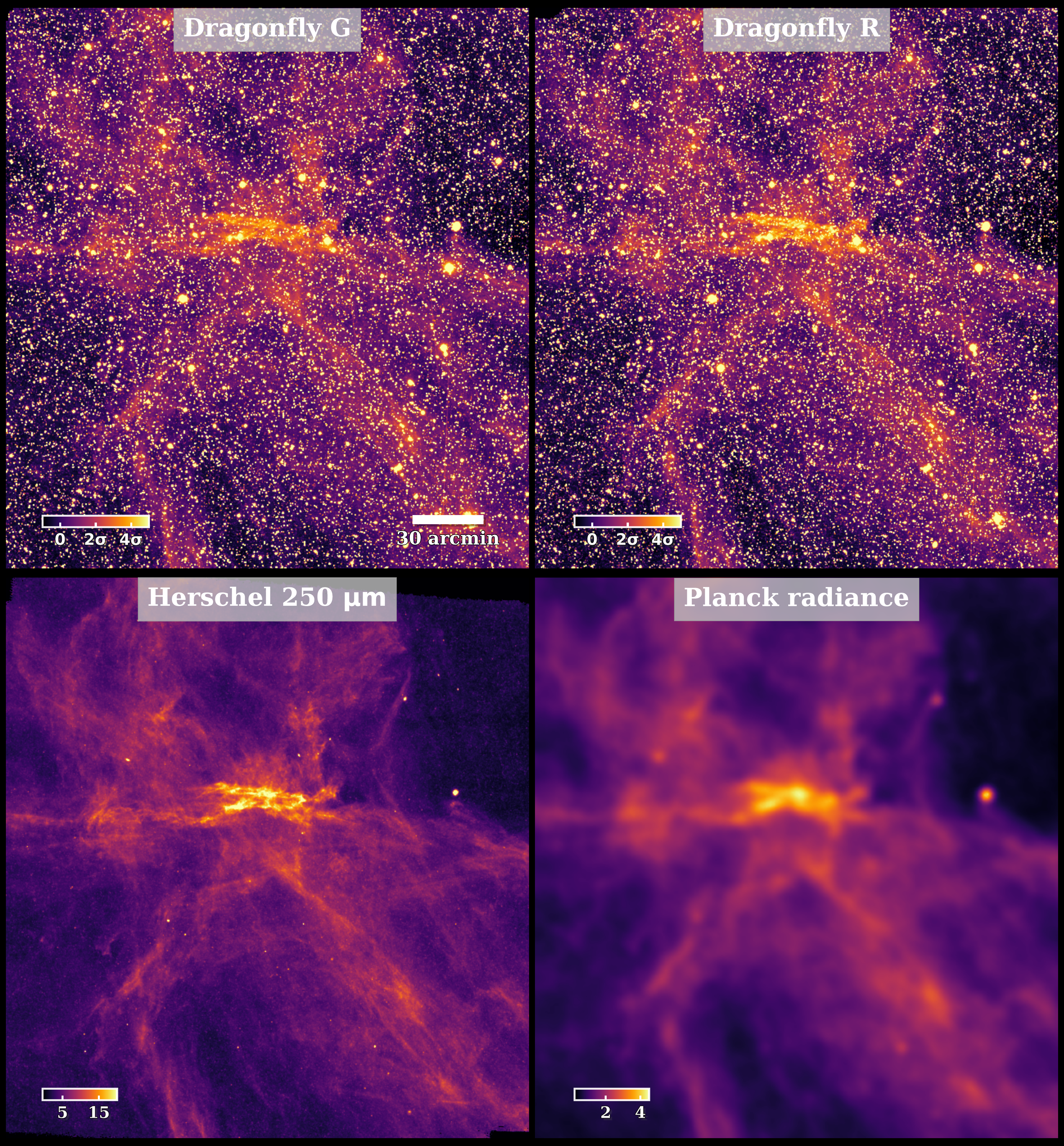}}
  \caption{\textit{Top left} and \textit{right}: Dragonfly \textit{g} and \textit{r}-band mosaic image of the Spider field. The field-of-view is $4.2^\circ\times4.5^\circ$. The colorbar is in units of {the Poisson noise of the stacked images. The Dragonfly seeing FWHM is $\sim 5 \arcsec$.} \textit{Bottom left}: Herschel SPIRE 250 micron image. The Herschel beam FWHM is 18$\arcsec$. The pixel values are in [$\rm{Myr{\,}sr^{-1}}$]. \textit{Bottom right}: Planck dust radiance map in units of [$10^{-7}{\,}\rm{W{\,}m^{-2}{\,}sr^{-1}}$]. The Planck beam FWHM is $5 \arcmin$. By visual comparison, the distribution of Galactic cirrus emission in the Spider field obtained by Dragonfly is in correspondence with that in Herschel and Planck, indicating a good performance on the preservation of low surface brightness emission.}
\label{fig:Spider_coadd}
\end{figure*}

Figures \ref{fig:Spider_rgb}--\ref{fig:uw1787_coadd} show the results obtained by putting the ideas in the previous three sections together and applying them to our example datasets: {The background models obtained in Sec. \ref{sec:sky_model} are subtracted from individual frames. The sky-subtracted images then undergo the wide-angle PSF assessment described in Sec. \ref{sec:wide_PSF}. After that, for the Spider fields, the GPR modeling demonstrated in Sec. \ref{sec:GP_model} is run to reject frames with inconsistent backgrounds.} The mosaic of the Spider field is displayed and compared with Herschel FIR data and Planck dust maps in Figure~\ref{fig:Spider_rgb} and Figure~\ref{fig:Spider_coadd}. The coadded stack of UW1787 is shown in Figure~\ref{fig:uw1787_rgb} and Figure~\ref{fig:uw1787_coadd} and compared with IRAS IR data and Planck dust map. Appendix~\ref{sec:coadd} describes how individual frames were combined into deep coadds for both fields and how the tiled mosaic for the Spider field was constructed. 

\subsection{Mosaic of the Spider Field} \label{sec:Spider_mosaic}
The final \textit{g}+\textit{r} mosaic of the full Spider field is displayed in Figure~\ref{fig:Spider_rgb} as an example data product. 
A comparison of the visible wavelength data from the Spider field with Herschel 250 micron data\footnote{The Herschel data were obtained by the Spectral and Photometric Imaging Receiver (SPIRE) instrument and retrieved from the online Herschel Science Archive (HSA): http://archives.esac.esa.int/hsa/whsa/.} and a Planck radiance map is shown in Figure \ref{fig:Spider_coadd}. The optical images obtained by Dragonfly show excellent correlation with the Herschel 250 micron emission and Planck dust radiance. In particular, the northwest corners of data in different wavelengths consistently exhibit an absence of cirrus emission. This is not the case with standard sky subtraction on single frames (the `poly' sky modeling in Fig.~\ref{fig:Planck_sky_1} and Fig.~\ref{fig:Planck_sky_2}), where we found an excess of light because the sky model was biased due to cirrus emission, and the attempt to account for the variation around field edges would oversubtract the signal there. This image reinforces the point in Section~\ref{sec:sky_model} that since the same dust population that emits thermal emission in sub-mm to FIR and scatters starlight in optical, visible wavelength data can be used to map galactic foregrounds traced at longer wavelengths.
In a future paper, we will explore in greater depth the correlation between Galactic cirrus emission at wavelengths from optical through the FIR and out to the sub-mm. We will also show how ancillary data can be used to model galactic cirrus, and demonstrate how these ideas can be used to improve the detectability of low surface brightness sources obscured by cirrus emission.

\subsection{Coadd of UW1787} \label{sec:uw_coadd}


\begin{figure*}[!htbp]
\centering
  \resizebox{0.6\hsize}{!}{\includegraphics{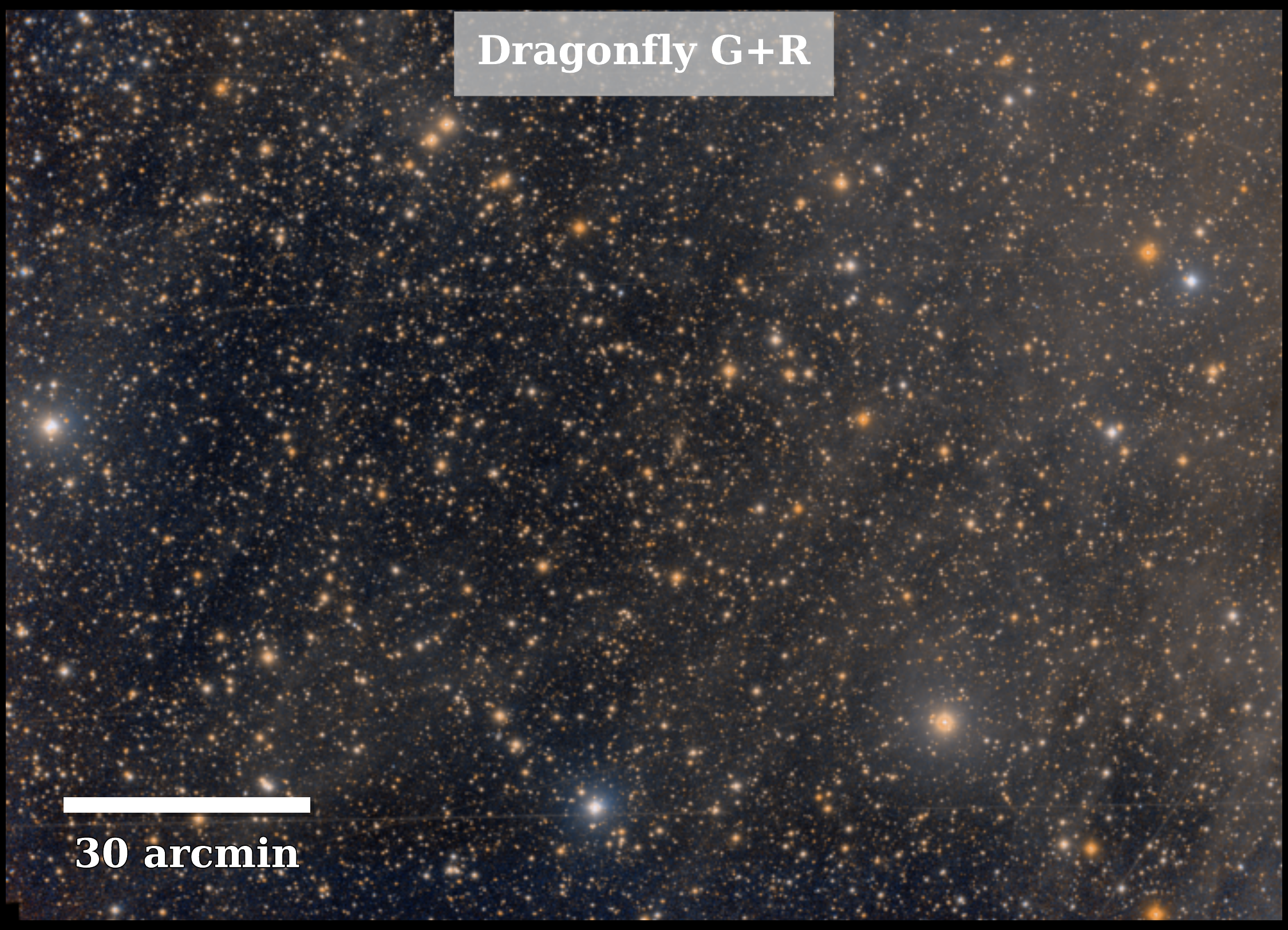}}
  \caption{Dragonfly \textit{g}+\textit{r} mosaic RGB image of UW1787, a typical field affected by high-galactic latitude cirrus in the Dragonfly Ultrawide survey. The G channel of the RGB is the average of \textit{g} and \textit{r} band data. The field-of-view is $2.8^\circ \times 2^\circ$.}
\label{fig:uw1787_rgb}
\end{figure*}

\begin{figure*}[!htbp]
\centering
  \resizebox{0.6\hsize}{!}{\includegraphics{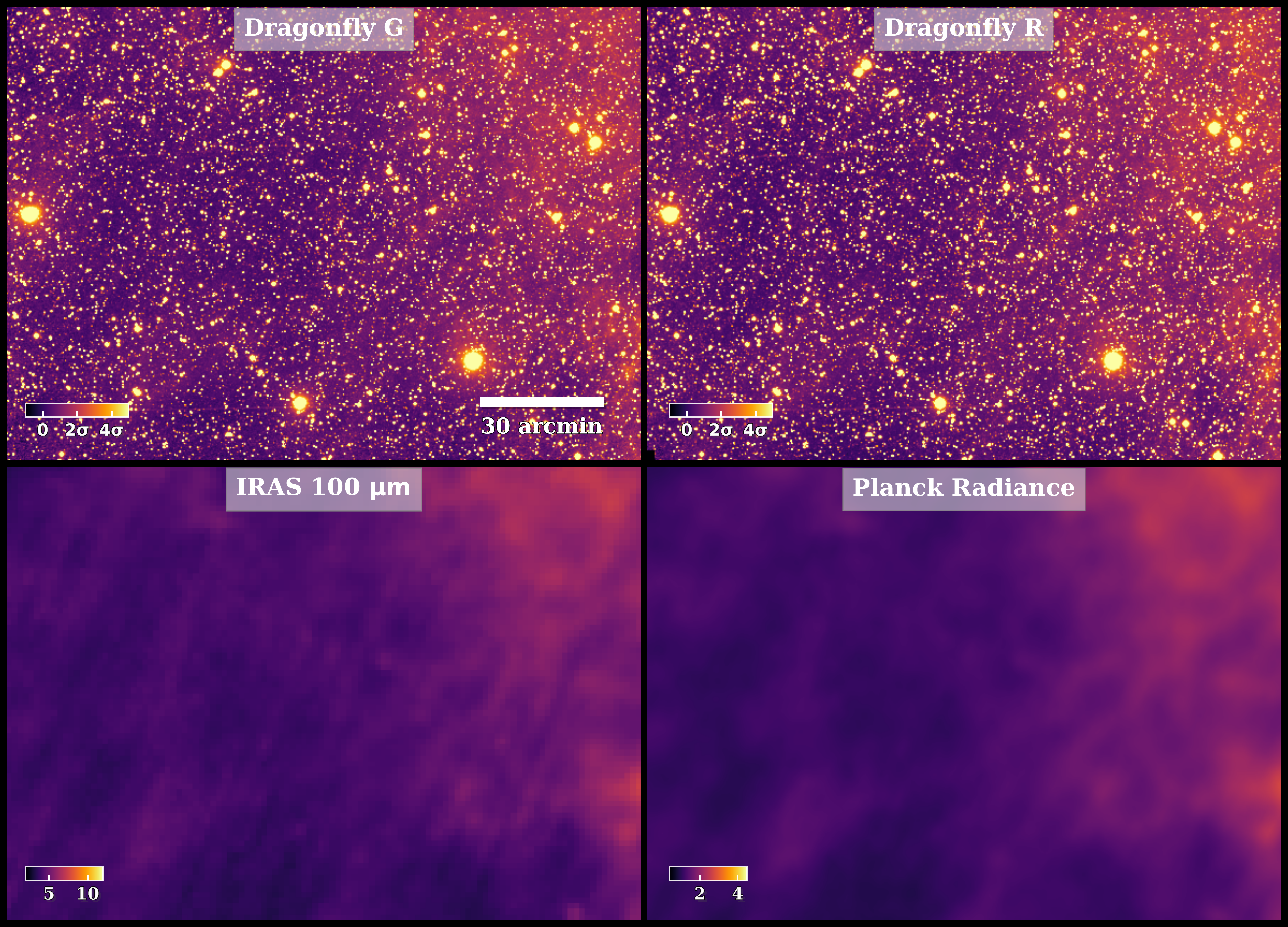}}
  \caption{\textit{Upper left} and \textit{right}: Dragonfly \textit{g} and \textit{r}-band coadd of UW1787. The field-of-view is $2.8^\circ \times 2^\circ$. The colorbar is in {units of the Poisson noise of the stacked images. The Dragonfly seeing FWHM is $\sim 5 \arcsec$.} \textit{Bottom left}: IRAS 100 micron image. The beam size is $4\arcmin$. The pixel values are in [$\rm{Myr{\,}sr^{-1}}$]. \textit{Bottom right}:  Planck dust radiance map in units of [$10^{-7}{\,}\rm{W{\,}m^{-2}{\,}sr^{-1}}$]. The Planck beam size is 5$\arcmin$.}
\label{fig:uw1787_coadd}
\end{figure*}

An RGB mosaic of UW1787 is displayed in Figure~\ref{fig:uw1787_rgb} and $g$-band and $r$-band images of this field are shown in Figure~\ref{fig:uw1787_coadd}. In this case, the cirrus emission is sufficiently weak such that normal sigma-clipping provides good fidelity, so we did not apply the GPR sky modeling technique presented in Section~\ref{sec:GP_model}, which assumes strong spatial correlations in the underlying signal. Because no Herschel data is available at this field's sky position, a 100-micron image retrieved from the IRAS Sky Survey Atlas (ISSA) Image Server\footnote{https://irsa.ipac.caltech.edu/applications/IRAS/ISSA/} is shown in Figure~\ref{fig:uw1787_coadd} for comparison. The Planck dust radiance map used for sky modeling is also shown in this figure. The techniques outlined in this paper reveal cirrus emission on the west side of the field, which would likely be suppressed if one adopts a conventional background modeling (as described in Sec.~\ref{sec:conventional_sky}). The cirrus emission that is present in the Dragonfly data exhibits an excellent match to the Planck data. This example demonstrates that our methods are able to preserve cirrus emission in the background of a typical field from a wide-field deep imaging survey.

\section{Discussion} \label{sec:discussion}

\subsection{Using Supplementary Data for Sky Modeling of Imaging Surveys with Telescopes Large and Small} \label{sec:discussion_surveys}

All-sky FIR/sub-mm dust emission data from Planck make it feasible to apply the techniques described in this paper to every position on the sky. However, the large beam size of the Planck satellite (with a FWHM of $\sim5\arcmin$) is a major challenge that prevents its synergy with optical data from typical deep wide-field imaging surveys obtained by large telescopes. Dragonfly data, suitably processed using the techniques presented here to preserve the dust emission, provide a path forward. Dragonfly's resolution is $\sim$50 times better than that of Planck, and these data are a bridge to images from large ground-based telescopes on excellent sites (such as CFHT and the 4 m Blanco Telescope), which in turn have a resolution $\sim$10 times better than Dragonfly, although they rely on mosaiced CCD sensors, which present formidable challenges to the preservation of the sky background on large scales. Combining the two datasets would result in the best of both worlds, by preserving the best ground-based resolution on images with unbiased sky backgrounds. This idea is one of the central reasons why we are undertaking the Dragonfly Wide Field Survey (\citealt{2020ApJ...894..119D}, \citealt{2021ApJ...909...74M}) and the Dragonfly Ultrawide Survey.


It is certainly possible (and probably a good idea) to use other supplementary long wavelength data, such as Herschel images, where available. In fact, a multi-wavelength synergy including all bands available is likely to provide the most robust sky model in terms of the preservation of the dust component. However, for the foreseeable future, Planck provides one of the best datasets to underpin the techniques described in this paper, because of its long wavelength all-sky coverage. Data from the IRAS satellite, with its nearly all-sky coverage, could also be used to supply priors; though, we caution that only its long wavelength channels should be used. In the near-infrared, emission comes mainly from small dust grains (e.g., \citealt{2001ApJ...554..778L}), which is not the exact same population that scatters the interstellar radiation field (ISRF) at visible wavelengths. Therefore, one should exercise extra caution before applying NIR data to model the background sky of visible wavelength data. Similar principles apply to Wide-field Infrared Survey Explorer (WISE), which has a much better resolution of 12$\arcsec$ at 22 $\mu m$, but requires further investigation to be used wisely. Long-wavelength channels of IRAS data, however, have similarly large beam sizes as Planck. In the absence of models, another (arguably more conservative) approach would be to apply FIR/sub-mm priors to optical imaging data just to flag regions 
where dust emission is above a certain threshold. For example, \cite{2019ApJS..240....1Z, 2021ApJS..257...60Z} used WISE, Planck, and IRAS maps to screen spurious UDG candidates. The authors found that, in cirrus-rich sky areas, up to 70\% of measurements could be affected by cirrus contamination.

The data from wide-field telescopes optimized for low surface brightness imaging can play an important role in calibrating other datasets. 
Future wide-field imaging surveys, such as the planned Legacy Survey of Space and Time (LSST) with the Vera Rubin Telescope, are likely to benefit from the Dragonfly Ultrawide Survey, which uses the techniques presented in this paper. 
Unlike the FIR/sub-mm sky, the optical low surface brightness sky background contains many sources other than the Galactic cirrus, including unresolved point sources and scattered light from bright stars and galaxies. We have developed post-processing techniques (\citealt{2020PASP..132g4503V}, \citealt{2022ApJ...925..219L}) to preserve low-frequency power on images, while making them maximally useful for investigations of higher-frequency structures. Obtaining clean, well-calibrated Galactic cirrus backgrounds from Dragonfly data is an ongoing subject of research and our progress will continue to be reported in future papers.

\subsection{Radiance or Tau? The Choice of Dust Surrogate in the Planck Dust Model} \label{sec:discussion_dust}

One of the decisions to be made in technique (1) (Section \ref{sec:sky_model}) is what IR/sub-mm product to use as a surrogate for the large-scale low-order spatial distribution and intensity of the scattered light cirrus seen in the optical. Both radiance and optical depth maps may be available, and while results obtained with either will be similar, they will not be identical. The best choice can be determined from physical principles relating the factors to be considered, as described below for faint high latitude cirrus that is optically thin at visible wavelengths.

A reasonable starting point, from what is known about the size distribution of dust particles from dust models (e.g., \citealt{2001ApJ...554..778L}, \citealt{2011A&A...525A.103C}), is that the scattering is dominated by the larger dust particles containing most of the mass, and it is these same particles that are large enough to be in thermal equilibrium with the ISRF. Both theoretically and empirically, these particles equilibrate at temperatures around 20 K, and the thermal dust emission is therefore in the sub-mm. This is the radiation described by the Planck all-sky dust model \citep{planck2013-p06}.
In that model, the SED in the sub-mm is modelled as an MBB law with three parameters: optical depth $\tau$, temperature $T$, and spectral index of the dust opacity, $\beta$ (the opacity $\kappa_\nu = \kappa_0 (\nu/\nu_0)^\beta$, in units [$cm^2 \cdot g^{-1} \cdot m^{-1}$], is the dust emission cross section per unit mass).

Following \citet{planck2013-p06b}, the intensity of the thermal radiation, $I_\nu$, is given by
\begin{equation}
\label{eq:inten}
I_\nu = \tau_\nu B_\nu(T) \,
\end{equation}
with
\begin{equation}
\label{eq:tau}
\tau_\nu = \kappa_\nu  M_{{\mathrm dust}}\, ,
\end{equation}
where $M_{{\mathrm dust}}$ is the mass column density of dust, and $B_\nu(T)$ is the Planck function.
The radiance $\cal{R}$ is
\begin{equation}
\label{eq:rad}
\mathcal{R} = \int I_\nu d\nu = \kappa_0 M_{{\mathrm dust}} \frac{\sigma_{{\mathrm S}}}{\pi} T^4 \left(\frac{k_BT}{h\nu_0}\right)^\beta \frac{\Gamma(4+\beta)\zeta(4+\beta)}{\Gamma(4)\zeta(4)} \, ,
\end{equation}
where $\sigma_{{\mathrm S}}$ is the Stefan-Boltzmann constant, $\Gamma$ and $\zeta$ are the Gamma and Riemann-zeta functions, respectively, and $h$ and $k_B$ are the Planck and Boltzmann constants, respectively.  If (i) $\kappa_0$ and $\beta$ are constant across the field, so is $T$, and therefore, $\cal{R}$ is proportional to (ii) $M_{{\mathrm dust}}$.

The radiance $\cal{R}$ is equal to the integral (over the size distribution and frequency) of the energy absorbed, which is proportional to the integral of the product of (iii) the number column density $N_{\rm dust}$, (iv) the (relative) size distribution of the larger dust particles, (v) the frequency dependent intensity of the ISRF, and (vi) the size and frequency dependent absorption cross section.

The total light scattered at a given frequency is the product of (iii), (iv), and (vii) the size and frequency dependent scattering cross section, integrated over the size distribution but not frequency. In addition, the intensity of scattered light $I_{{\mathrm scat}}$ received by an observer is affected by the combination of (viii) the frequency dependent directional anisotropy of the ISRF and (ix) the size and frequency dependent anisotropic scattering phase function.

In applying technique (1), we require that the surrogate for the scattered light is valid at least over the limited spatial field being analyzed (not the whole sky), to be multiplied by a spatially independent constant of proportionality, $\lambda$, which is unknown.  Local uniformity seems plausible for the ISRF-related factors (v) and (viii) and for the dust population related factors (ii), (iii), (iv), (vii), and (ix), so that both $\cal{R}$ and $I_{{\mathrm scat}}$ are modulated spatially in the same way by the column density.  However, in different parts of the sky, spatial changes in any of these factors, for example the combination of (viii) and (ix), could change the constant of proportionality.

Suppose that $\cal{R}$ simplifies to being a surrogate for the column density. Are there alternates? From Equation \ref{eq:tau}, if (i) $\kappa_0$ is constant across the field, then $\tau_{\nu_0}$ as a surrogate of $M_{{\mathrm dust}}$ would be acceptable.  However, in the case of the Spider field, there is evidence in the Planck dust model that $T$ is lower with increasing $\cal{R}$ (or $\tau_{\nu_0}$).
Lower temperature can be explained by higher $\kappa_0$, because then the dust is able to radiate more efficiently and equilibrate at a lower temperature for the same energy absorbed.\footnote{There is a complicating effect of small changes in the spectral index of the opacity as well, which we ignore just to simplify this exposition.}
In thermal equilibrium, the product $\kappa_0 T^{4+\beta}$ is preserved. Therefore, $\tau_{\nu_0}$ is modulated not only by $M_{{\mathrm dust}}$ but also as $1/T^{4+\beta}$. Then, as observed in the Planck dust model solution, $\tau_{\nu_0}$ increases faster than linearly with $\cal{R}$.

The observed specific intensity needs to be matched at each passband by the $I_\nu$ of MBB model (Equation \ref{eq:inten}). Where spatially $T$ is lower, the fitted MBB parameter $\tau_{\nu_0}$ is larger as described, but this is reduced by $B(\nu, T)$ being lower. In the Rayleigh-Jeans limit, $B(\nu, T)$ is reduced in proportion to just $T$, but closer to the peak of the SED, the reduction is greater, roughly compensating the increase in $\tau_{\nu_0}$, so that $I_\nu$ for that range of $\nu$ is roughly proportional to column density and $\cal{R}$.  Past the peak, the reduction is even higher, so the $I_\nu$ rises more slowly than linearly with $\cal{R}$.

While this provides some guidance based on physical considerations, it does not span all possibilities.  Briefly, caveats include the following. If $\kappa_0$ were changing spatially, would the optical cross sections remain unaffected? Even if there are changes in $T$ so that $\tau_{\nu_0}$ does not track $\cal{R}$, that is not the only consideration; the spatial morphology of the changes is important to the derivation of the polynomials too. Not all fields reveal temperature changes; even for Spider, MBB solutions using Herschel data \citep{singhmartin2022} do not show as systematic changes of $T$ with $\cal{R}$, possibly because of the different distributions of passbands relative to the SED peak or different weightings in the fit. The map of $\cal{R}$ is less noisy than the map of $\tau_{\nu_0}$; however, the fine-scale noise should not impact the low-order large-scale polynomials being fit.

\section{Summary} \label{sec:summary}

Unbiased sky background modeling is crucial for the detection and measurement of diffuse emission in deep wide-field imaging. Obtaining an unbiased sky background model requires two things: (1) removal/control of the large-scale time-varying components in the image, including (but not limited by) zodiacal light, scattered light from the wide-angle point spread function, flat-fielding errors, and camera artifacts; and (2) preservation of astronomically interesting large-scale signal, originating from a host of phenomena, including low surface brightness galaxies, galactic stellar halos, and Galactic cirrus. These requirements are hard to meet, because the diffuse emission of interest is nearly indistinguishable from the sky background at low surface brightness levels. Conventional sky modeling techniques for handling these problems tend to over-subtract backgrounds in the final data products.

In this paper, we describe three methods that can be applied to wide-field visible wavelength data to improve sky models. These include the following: 
\begin{enumerate}
\item Use of FIR/sub-mm dust {templates} in the background modeling {with two assumptions: (a) the underlying sky background has little power on small spatial scales, and (b) the Galactic cirrus in the field is optically thin on the scale of sky modeling}. Publicly available Planck dust maps are particularly useful sources of information for the application of this technique. 
\item Filtering frames with abnormal wide-angle PSFs to control the amount of scattered light that originates from the extended wings of bright stars. 
This reduces systematics caused by temporal variations in the wide-angle PSF, and increases the signal-to-noise levels of low surface brightness structures seen in deep image stacks.
\item An image combination method making use of pixel covariance, which is more robust and effective than sigma-clipping in controlling local background variations. Assuming the spatial correlation and time-invariability of the underlying extended diffuse emission, the background in multiple observations of multiple patches of the sky can be modelled with a Gaussian process in which pixel areas with large deviations, which indicate inconsistent backgrounds, are rejected or downweighted in the stacking process.
\end{enumerate}

The methods for background modeling presented in this paper are three ingredients in a recipe for data reduction used by the Dragonfly Telephoto Array. In the present paper, we illustrate the efficacy of this recipe using two example datasets. The first is a $\sim$20 deg$^2$ field centered upon the prominent Spider Galactic cirrus complex. The second example is UW1787, a typical field from the Dragonfly Ultrawide Survey affected by diffuse high Galactic latitude cirrus backgrounds. In both cases, 
final stacked images show significantly improved sky background systematics.
These techniques {will be incorporated into the future version} of the image reduction pipeline used by the Dragonfly Telephoto Array. {Our codes are publicly available for interested readers.}\footnote{https://github.com/DragonflyTelescope/DFCirrus}$^,$\footnote{https://github.com/DragonflyTelescope/DFHalo} By applying these techniques, data obtained from wide-field telescopes optimized for low surface brightness imaging can play an important role in calibrating the backgrounds in other surveys, such as the LSST to be undertaken by the Vera Rubin Telescope starting in 2024.

\begin{acknowledgments}
Q.L. is supported by an Ontario Trillium Award and an Ontario Graduate Scholarship. The research of R.G.A. and P.G.M. is supported by grants from the Natural Sciences and Engineering Research Council of Canada. The Dunlap Institute is funded through an endowment established by the David Dunlap family and the University of Toronto. This research has made use of data from the Planck Legacy Archive (PLA). The Planck Legacy Archive provides online access to all official data products generated by the Planck mission. This research has made use of data from the Legacy Surveys. 
This research has made use of the APASS database, located at the AAVSO website. Funding for APASS has been provided by the Robert Martin Ayers Sciences Fund. The authors thank the excellent and dedicated staff at the New Mexico Skies Observatory.

\end{acknowledgments}

\vspace{5mm}


\software{astropy \citep{2013A&A...558A..33A, 2018AJ....156..123A, 2022ApJ...935..167A}, numpy \citep{harris2020array}, scipy \citep{2020NatMe..17..261V}, matplotlib \citep{Hunter:2007},
          Source Extractor \citep{1996A&AS..117..393B}, reproject \citep{2020ascl.soft11023R}, photutils \citep{2016ascl.soft09011B}, scikit-image \citep{van2014scikit}, scikit-learn \citep{2011JMLR...12.2825P}
          }



\appendix

\section{Telescope, Observation, and Data Reduction}

\subsection{The Dragonfly Telephoto Array} \label{sec:telescope}
The Dragonfly Telephoto Array is a mosaic aperture telescope comprised of 48 Canon 400 mm $f/2.8$ IS II USM-L telephoto lenses. Dragonfly is optimized for low surface brightness imaging in its concept of design. Each individual lens is equipped with a Santa Barbara Imaging Group (SBIG) CCD camera with a field of view of 2.6$^\circ$ $\times$ 1.9$^\circ$ and a pixel scale of 2.85\arcsec/pix. It is equivalent to a 1.0 m $f/0.39$ refractor with a field of view of 2$^\circ$ $\times$ 3$^\circ$. With well-baffled all-refractive optics (key components of which have sub-wavelength nanostructure coatings) and no pupil obscuration, scattered light in Dragonfly's optical path is minimized for the observation of low surface brightness features. Half (24) of the cameras take images in the Sloan \textit{g}-band, and the other half take images in the Sloan \textit{r}-band. To eliminate camera-to-camera systematics, individual lens pointings are offset by small amounts relative to each other, and images are obtained with large dithers (from $1\arcmin$ to $15\arcmin$, depending on the target) to produce the final image coadd. We refer the reader to \cite{2014PASP..126...55A} for the general principles of the design and to \cite{2020ApJ...894..119D} for a description of the latest configuration of the array (which is growing) and for our current observing strategies.

\subsection{Data Acquisition \& Reduction} \label{sec:reduction}

In this work, we used images obtained by Dragonfly in 2021--2022 to illustrate reduction methods. For each field, following the observing strategy described in \citealt{2020ApJ...894..119D}, Dragonfly took 10 minute exposures and did a preliminary assessment of image quality based on the number, FWHM, ellipticity, and zero-points of detected sources, until a proposed number of frames (half in \textit{g}-band and half in \textit{r}-band) passed quality control cuts. {The mean FHWM of Dragonfly's PSF at New Mexico Skies under good conditions is $\approx 5\arcsec$.} Frames were taken with large dithers to reduce systematics. {The dithering angle was $5\arcmin$ for Spider fields and $1\arcmin$ for UW data.} The raw frames were then bias subtracted, dark subtracted, and flat-fielded using utilities in the Dragonfly data reduction pipeline \texttt{dfreduce} (see \citealt{2020ApJ...894..119D}). The flats were taken at the start and the end of the observing night and flats from adjacent nights were combined into master flats. Astrometric solutions were derived using the \texttt{astrometry.net} module (\citealt{2010AJ....139.1782L}). Pixel area maps were used to correct the distortion by the airmass difference across the field during this process.

In the standard Dragonfly data reduction pipeline, the sky background is subtracted with two iterations of polynomial fitting, with the second iteration occurring after masking sources and bad pixels identified in the first iteration. In the main text of the present paper, we illustrate the benefits of a different sky background treatment from that used in the standard reduction pipeline. The main reason for this is that, in cirrus-rich areas, a polynomial estimate of the sky background suffers from strong biases (details illustrated in Section~\ref{sec:sky_model}). Therefore, the techniques described in this paper are being incorporated into the next generation pipeline for the instrument.

\section{Coadding, Sky Matching, and Tiling} \label{sec:coadd}

The processed frames used to illustrate the techniques described in this paper were combined into deep coadds as described in Section~\ref{sec:mosaic}. We used a median combination to generate the coadded image, since this best removes satellite trails. The deviation maps generated by Gaussian process background modeling were used to reject pixel areas with large deviation ($>$1\%) relative to the mean pattern for fields with strong cirrus emission. The mosaic of the Spider field presented in this paper is constructed from coadds stacked following these criteria. We have also generated deep coadds combined with weighted averaging using the deviation maps.

Because the mean sky values are different among deep coadds, a small correction that shifts the deep coadds to a common sky level is necessary to tile them into a mosaic. The sky correction is done by calculating the mean sky offsets of coadd pairs in overlapping regions, which is similar to the \texttt{`globalmin+match'} algorithm in the \texttt{skymatch} utility of the \texttt{stsci.skypac} Python module. Prior to this, we do a zero-point correction to standardize the zero-point of the mosaic. The procedure is as follows: First, we calculate the zero-points of all coadd images using unsaturated stars between 14 to 16.5 mag cross-matched from the APASS star catalog (\citealt{2016yCat.2336....0H}). The images are scaled based on the differences of zero-points to the median value. Given a pair of overlapping images, we measure the local background (sky + signal) with a mode estimator using a box size of [64 x 64] pix$^2$. Compact sources are masked during the measurement. Only vertical or horizontal pairs are measured, i.e., images with diagonal overlaps are excluded. We take the median of the background differences of each pair as their sky offsets. Then, we equalize the sky offsets by taking the average value and setting the minimum value to 0. 

After shifting the coadd images into a common sky value, they are reprojected into the same grid using the Python package \texttt{reproject} and tiled into the mosaic. In the overlapping regions, the coadds are combined using an average combination weighted by the number of available frames. The field edges with low overlapping coverage are rejected in the combination. 

\section{Further Examples of Sky Modeling in the Spider Field} \label{sec:sky_model_spider_fig}

\begin{figure*}[!htbp]
    \centering
    \includegraphics[width=\textwidth]{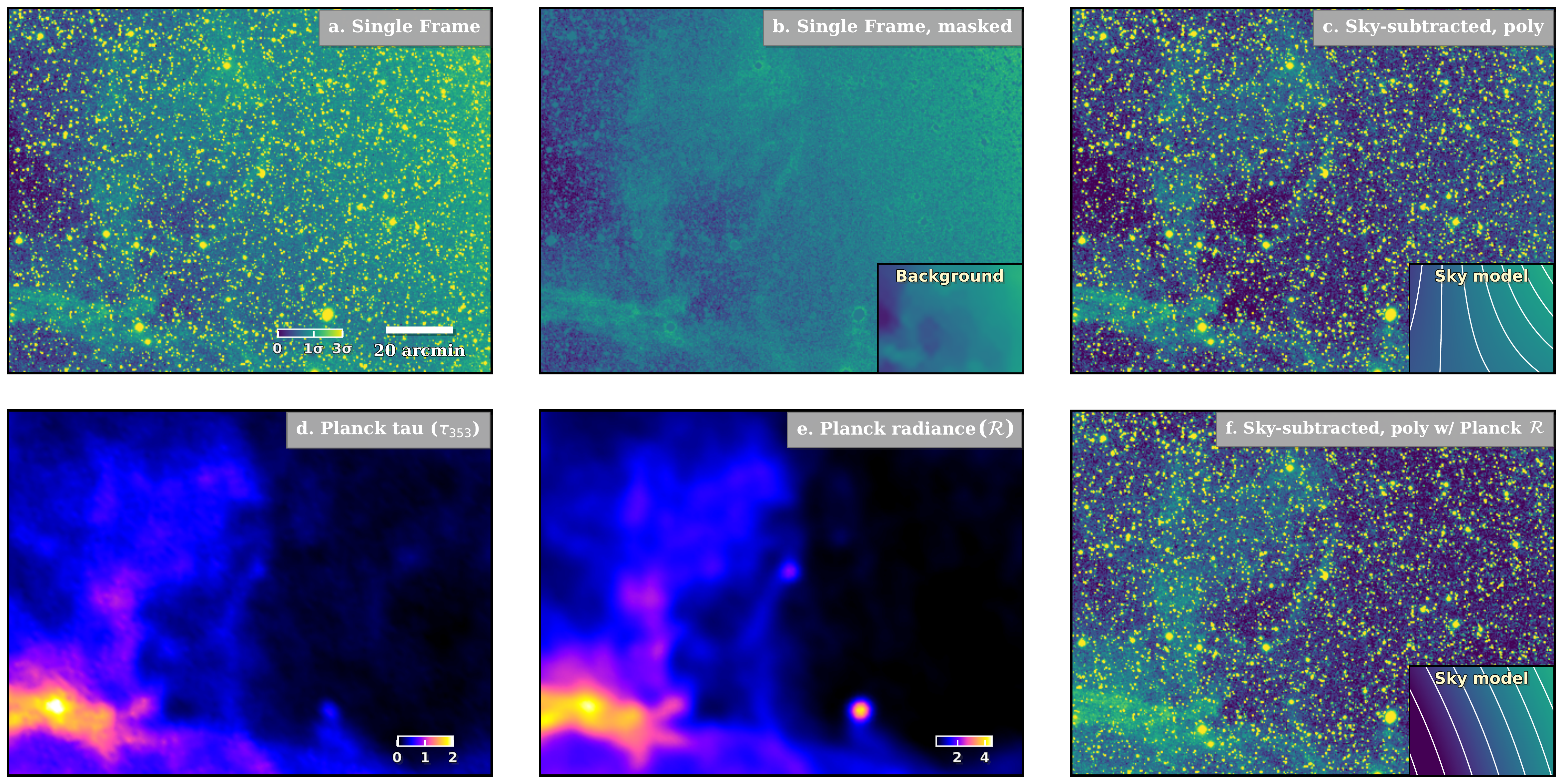}
    \caption{Comparison of sky modeling of an individual frame using a single polynomial model and that with Planck dust model. The image is a $2.6^{\circ} \times 2^{\circ}$ 10-min exposure from a single camera of Dragonfly. The layout is the same as Fig.~\ref{fig:Planck_sky_1}, but shows a different frame of the Spider field, which presents stronger large-scale sky variation and a void of dust emission. The sky pattern near the right edge of the image is under-subtracted and cirrus on the lower left is suppressed by a single polynomial sky model, compared to the result {with the incorporation} of Planck templates.}
    \label{fig:Planck_sky_2}
\end{figure*}

\begin{figure*}[!htbp]
    \centering
    \includegraphics[width=\textwidth]{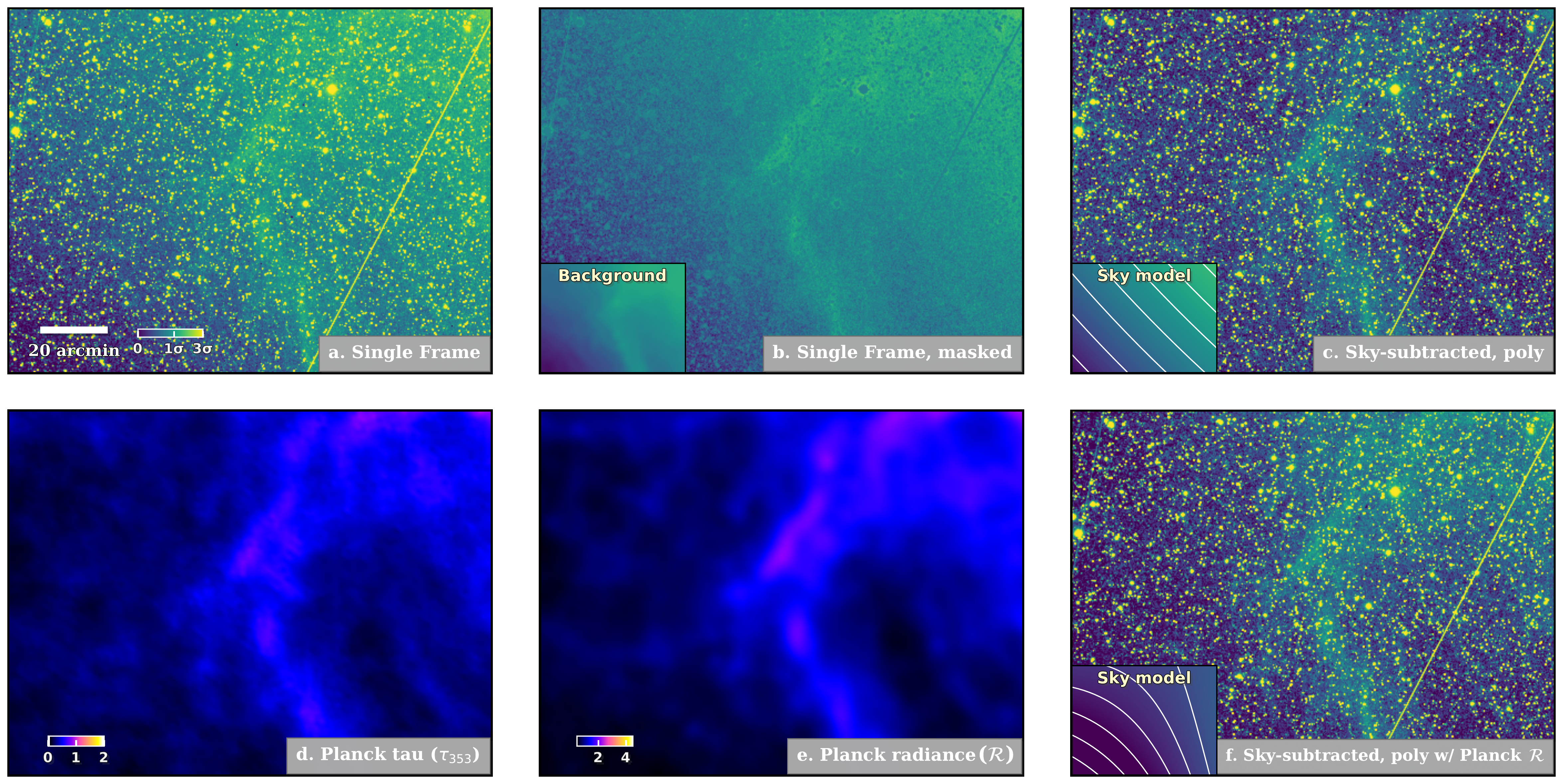}
    \caption{Comparison of sky modeling of an individual frame using a single polynomial model and that with Planck dust model. The image is a $2.6^{\circ} \times 2^{\circ}$ 10 minute exposure from a single camera of Dragonfly. A satellite trail presents in the image. The layout is the same as Fig.~\ref{fig:Planck_sky_1} but shows a different frame of the Spider field with diffuse cirrus structures coupled with the large-scale sky. The sky pattern near the left edge of the field is under-subtracted, and cirrus emission on the upper right is suppressed by a single polynomial sky model, compared to the result  {with the incorporation} of Planck templates.}
    \label{fig:Planck_sky_3}
\end{figure*}

In Section \ref{sec:sky_model}, Fig.~\ref{fig:Planck_sky_1} presents an example of the application of sky modeling based on dust models using one single frame that covers the central bright region of the Spider field. The morphology in the Spider field is complex; however, it turns out that this sky modeling method applies well throughout the field. Figures \ref{fig:Planck_sky_2} and \ref{fig:Planck_sky_3} present more examples of frames with different cirrus structures. Their layouts are the same as Fig.~\ref{fig:Planck_sky_1}.
Fig.~\ref{fig:Planck_sky_2} shows an example similar to Fig.~\ref{fig:Planck_sky_1} with prominent cirrus structures around field edges, but with a stronger sky variation and a large area where dust emission is absent. In panel (c), the conventional sky model is biased by the cirrus on the lower left of the image, and it fails to delineate the variation around the right edge of the field, leading to an over-subtraction of the cirrus and residual light of the underlying sky on the right. Fig.~\ref{fig:Planck_sky_3} shows an example of more diffuse cirrus structures that are strongly decoupled with the large-scale sky background. Panel (c) of Fig.~\ref{fig:Planck_sky_3} shows that the sky model dramatically suppresses the diffuse cirrus structures, while leaving residual light near the left edge of the field. The sky modeling is significantly improved in both cases, as shown in panel~(f) of Figures~\ref{fig:Planck_sky_2} and \ref{fig:Planck_sky_3}. In summary, the sky model described in Sec. \ref{sec:sky_model} shows abilities to handle a variety of spatial distributions of cirrus in the FoV.

\section{Example of Wide-angle PSF Assessment in UW1787} \label{sec:wide_psf_uw_fig}

\begin{figure*}[!htbp]
\centering
  \resizebox{0.98\hsize}{!}{\includegraphics{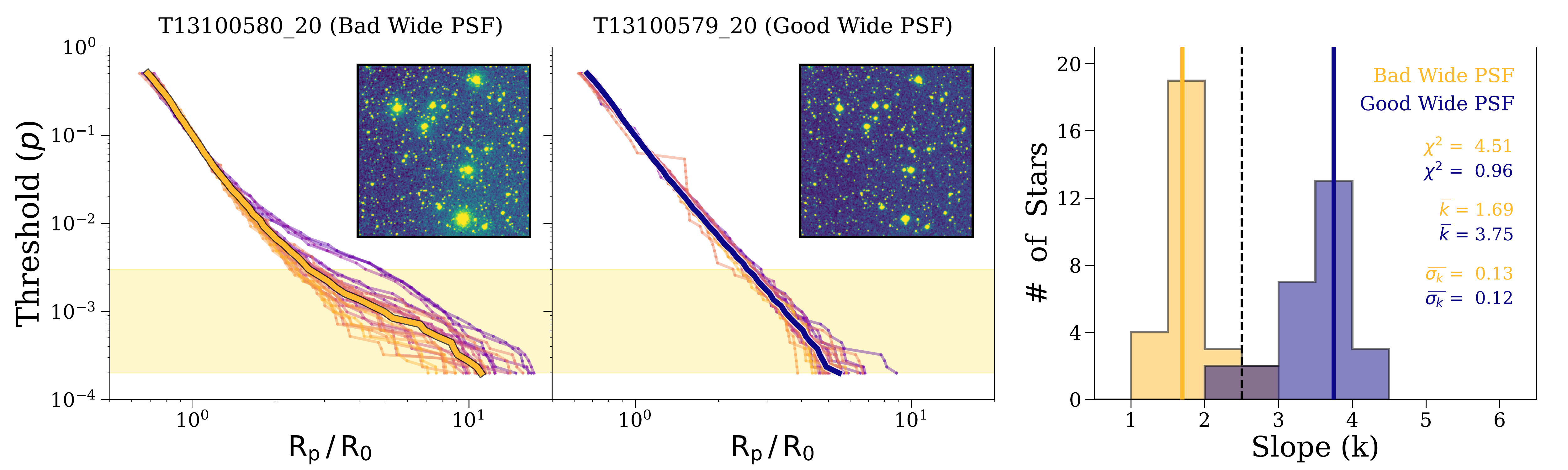}}
  \caption{Wide-angle PSF assessment of two examples frames of UW1787 based on absolute metrics. \textit{Left} and \textit{middle}: threshold profiles extracted from a single frame flagged with bad and good wide-angle PSF. The yellow shaded areas indicate the threshold range in which the slopes are measured. The postage stamps show 40$\arcmin$ $\times$ 40$\arcmin$ cutouts of the frames. \textit{Right}: histograms of slopes measured from profiles of individual stars of the two frames (orange: bad; blue: good). The black dashed line shows the demarcation line (bad/good wide-PSF) based on median slopes.} 
\label{fig:halo_frame_uw1787}
\end{figure*}

In Section \ref{sec:wide_PSF}, we present a method of assessing the goodness of wide-angle PSF. Figure~\ref{fig:halo_frame} illustrates the assessment based on absolute metrics using example frames of the Spider field. Here we show further examples in Figure~\ref{fig:halo_frame_uw1787} using two frames of the UW1787 field. As in Figure~\ref{fig:halo_frame}, the two frames were taken at the same time but with different cameras. The left and right panels of Figure~\ref{fig:halo_frame_uw1787} show profiles extracted from a frame flagged classified as bad and good, which is also revealed in the postage cutouts. The right histograms show the distributions of slopes measured on individual stars and the measured metrics described in Section \ref{sec:metrics_wide_psf} used for the assessment of wide-angle PSF.

\bibliography{bibtex, Planck_bib}{}
\bibliographystyle{aasjournal}



\end{document}